\documentclass[12pt,journal,draftcls,a4paper,onecolumn]{IEEEtran}
\usepackage{dsfont}

\usepackage{subeqnarray}
\usepackage{cleveref}
\usepackage{amssymb}
\usepackage{amsfonts}
\usepackage{amsmath}
\usepackage{graphicx}
\usepackage{amsmath}
\usepackage[all,poly]{xy}
\usepackage{multirow}
\usepackage{algorithm}
\usepackage{algorithmic}
\usepackage{color}
\usepackage[noadjust]{cite}

\title{Residual component analysis of hyperspectral images -- Application to joint nonlinear \\unmixing and nonlinearity detection}
\author{}
\author{Yoann Altmann, Nicolas Dobigeon, Steve
McLaughlin and Jean-Yves Tourneret
\thanks{Part of this work was supported by the Direction G\'en\'erale de l'armement, French Ministry of Defence, the \textsc{Madonna} project funded by INP Toulouse
and the \textsc{Hypanema} ANR Project n$^\circ$ANR-12-BS03-003.}
\thanks{Yoann Altmann, Nicolas Dobigeon and Jean-Yves Tourneret are with University of Toulouse, IRIT/INP-ENSEEIHT, 2 rue Charles Camichel, BP 7122, 31071 Toulouse cedex 7,
France (e-mail:
\{Yoann.Altmann, Nicolas.Dobigeon, Jean-Yves.Tourneret\}@enseeiht.fr).}\\
\thanks{Steve McLaughlin is with School of Engineering and Physical Sciences, Heriot-Watt University,
U.K. (email: s.mclaughlin@hw.ac.uk).}}
\newcommand{\bc}{\boldsymbol{c}}

\newcommand{\bs}{\boldsymbol{s}}

\newcommand{\bz}{\boldsymbol{z}}

\newcommand{\bK}{\boldsymbol{K}}
\newcommand{\bQ}{\boldsymbol{Q}}

\newcommand{\bX}{{\boldsymbol X}}

\newcommand{\bsigma}{\boldsymbol{\sigma}}
\newcommand{\bSigma}{\boldsymbol{\Sigma}}
\newcommand{\bPsi}{{\boldsymbol \Psi}}

\newcommand{\bphi}{{\boldsymbol \phi}}
\newcommand{\bPhi}{{\boldsymbol \Phi}}

\newcommand{\blue}{\textcolor{blue} }
\def\calC{{\mathcal{C}}}

\newcommand{\Vpix}[1]{\mathbf{y}_{#1}}

\newcommand{\MATpix}{\mathbf{Y}}
\newcommand{\pix}[2]{y_{#1,#2}}

\newcommand{\Vpixels}{\mathbf{y}}

\newcommand{\nopix}{n}

\newcommand{\nbband}{L}

\newcommand{\nbmat}{R}
\newcommand{\nomat}{r}

\newcommand{\MATmat}{{\mathbf M}}
\newcommand{\Vmat}[1]{{\mathbf m}_{#1}}
\newcommand{\mat}[2]{m_{#1,#2}}

\newcommand{\MATabond}{{\bold A}}
\newcommand{\Vcoeff}[1]{{\bold c}_{#1}}

\newcommand{\MATcoeff}{{\bold C}}

\newcommand{\abond}[2]{{a}_{#1,#2}}
\newcommand{\Vabond}[1]{{\boldsymbol{a}}_{#1}}

\newcommand{\Vabonds}{{\boldsymbol{a}}}

\newcommand{\Vnoise}{{\mathbf e}}

\newcommand{\noisevar}{\sigma^2}

\newcommand{\paramvect}{\boldsymbol{\theta}}

\newcommand{\Simplex}{\mathcal{S}}

\newcommand{\R}{\mathds{R}}

\newcommand{\transp}{^T}

\newcommand{\etr}{\mathrm{etr}}

\newcommand{\diag}[1]{\textrm{diag}\left(#1\right)}
\newcommand{\Ndistr}[1]{\mathcal{N}\left(#1\right)}

\newcommand{\norm}[1]{\left\|#1\right\|}

\newcommand{\Vzero}{\boldsymbol{0}}
\newcommand{\Indicfun}[2]{\textbf{1}_{#1}\left(#2\right)}

\newcounter{algo}
\renewcommand{\thealgo}{\arabic{algo}}

\begin{document}
\maketitle

\begin{abstract}
This paper presents a nonlinear mixing model for joint hyperspectral
image unmixing and nonlinearity detection. The proposed model
assumes that the pixel reflectances are linear combinations of known
pure spectral components corrupted by an additional nonlinear term,
affecting the endmembers and contaminated by an additive Gaussian
noise. A Markov random field is considered for nonlinearity
detection based on the spatial structure of the nonlinear terms. The
observed image is segmented into regions where nonlinear terms, if
present, share similar statistical properties. A Bayesian algorithm
is proposed to estimate the parameters involved in the model
yielding a joint nonlinear unmixing and nonlinearity detection
algorithm. The performance of the proposed strategy is first
evaluated on synthetic data. Simulations conducted with real data
show the accuracy of the proposed unmixing and nonlinearity
detection strategy for the analysis of hyperspectral images.
\end{abstract}

%\begin{center} \bfseries EDICS: COI-RRG (Radar Imaging, Remote Sensing, and Geophysical Imaging)
%\end{center}

\section{Introduction}
Spectral unmixing (SU) of hyperspectral images has attracted growing
interest over the last few decades. It consists of distinguishing
the materials and quantifying their proportions in each pixel of the
observed image. This blind source separation problem has been widely
studied for the applications where pixel reflectances are linear
combinations of pure component spectra
\cite{Craig1994,Heinz2001,Eches2010a,Miao2007a,Yang2011}. However,
as explained in \cite{Keshava2002,Bioucas2012}, the linear mixing
model (LMM) can be inappropriate for some hyperspectral images, such
as those containing sand, trees or vegetation areas. Nonlinear
mixing models (NLMMs) provide an interesting alternative to
overcoming the inherent limitations of the LMM. They have been
proposed in the hyperspectral image literature and can be divided
into two main classes \cite{Dobigeon2013spmag}.

The first class of NLMMs consists of physical models based on the
nature of the environment. These models include the bidirectional
reflectance based model proposed in \cite{Hapke1981} for intimate
mixtures associated with sand-like materials and the bilinear models
recently studied in
\cite{Somers2009,Nascimento2009,Fan2009,Halimi2010} to account for
scattering effects mainly observed in vegetation and urban areas.
The second class of NLMMs contains more flexible models allowing for
different kinds of nonlinearities to be approximated. These flexible
models are constructed from neural networks
\cite{Guilfoyle2001,Altmann2011IGARSS}, kernels
\cite{Chen2012,Altmann2013}, or post-nonlinear transformations
\cite{Altmann2012a}.

Unfortunately, developing nonlinear unmixing strategies and refining
mixing models usually implies a high computational cost. While
consideration of nonlinear effects can be relevant in specific
areas, the LMM is often sufficient for approximating the actual
mixing models in some image pixels, for instance in homogeneous
regions. To reduce the complexity required to process an image, it
makes sense to distinguish in any image, linearly mixed pixels which
can be easily analyzed, from those nonlinearly mixed requiring
deeper analysis. Nonlinearity detection in hyperspectral images has
already been addressed in \cite{Han2008} to detect nonlinear areas
in observed scenes using surrogate data. In previous work, a
pixel-by-pixel nonlinearity detector based on a polynomial
post-nonlinear mixing model (PPNMM) was proposed and provided
interesting results \cite{Altmann2012b}. The detector in
\cite{Altmann2012b} follows a PPNMM-based SU procedure and uses the
statistical properties of the parameter estimator to subsequently
derive an accurate test statistic. Conversely, this paper proposes
to simultaneously achieve the SU and nonlinearity detection. Moreover, it was noted in \cite{Altmann2012b} that
the consideration of spatial structures in the image, already used
in \cite{Eches2011} for linear SU, can also be used to infer the
locations where nonlinear effects occur.

This paper presents a new supervised Bayesian algorithm for joint
nonlinear SU and nonlinearity detection. This algorithm is
supervised in the sense that the endmembers contained in the image
are assumed to be known (chosen from a spectral library or extracted
from the data by an endmember extraction algorithm (EEA)). This
algorithm is based on a nonlinear mixing model inspired from
residual component analysis (RCA) \cite{Kalaitzis2012}. In the
context of SU of hyperspectral images, the nonlinear effects are
modeled by additive perturbation terms characterized by Gaussian
processes (GPs). This allows the nonlinear terms to be marginalized,
yielding a flexible model depending only on the nonlinearity energies.
The hyperspectral image to be analyzed is
partitioned into homogeneous regions in which the nonlinearities
share the same GP. This algorithm relies on an implicit image
classification, modeled by labels whose spatial dependencies follow
a Potts-Markov random field. Consideration of two classes (linear
vs. nonlinear mixtures) would lead to binary detection maps.
However, this paper allows for nonlinearly mixed regions to be also
identified, based on the energy of the nonlinear effects. More
precisely, the proposed algorithm can identify regions with
different level of nonlinearity and characterized by different GPs.
Most SU algorithms assume additive, independent and identically
distributed (i.i.d.) noise sequences. However, based on
previous work conducted on real hyperspectral images, non i.i.d. noise vectors
are considered in this paper.

In the Bayesian framework, appropriate prior distributions are
chosen for the unknown parameters of the proposed RCA model, i.e.,
the mixing coefficients, the GP hyperparameters, the class labels and the
noise covariance matrix. The joint posterior distribution of these
parameters is then derived. However, the classical Bayesian
estimators cannot be easily computed from this joint posterior. To
alleviate this problem, a Markov chain Monte Carlo (MCMC) method is
used to generate samples according to the posterior of interest.
Finally, the generated samples are used to compute Bayesian
estimators as well as measures of uncertainties such as confidence
intervals.

The remaining paper is organized as follows. Section
\ref{sec:Problem} introduces the RCA model for hyperspectral image
analysis. Section \ref{sec:bayesian} presents the hierarchical
Bayesian model associated with the proposed RCA model and its
posterior distribution. The Metropolis-Within-Gibbs sampler used to
sample from the posterior of interest is detailed in Section
\ref{sec:Gibbs}. Some simulation results conducted on synthetic and
real data are shown and discussed in Sections \ref{sec:simu_synth}
and \ref{sec:simu_real}. Conclusions are finally reported in Section
\ref{sec:conclusion}.

\section{Problem formulation}

\label{sec:Problem} We consider a set of $N$ observed pixel spectra
$\Vpix{n} = [\pix{n}{1},\ldots,\pix{n}{\nbband}]\transp, n \in \left
\lbrace 1,\ldots,N \right \rbrace$ where $\nbband$ is the number of
spectral bands. Each of these spectra is defined as a linear
combination of $R$ known spectra $\Vmat{r}$, referred to as
endmembers, contaminated by an additional spectrum $\bphi_{n}$ and
additive noise
\begin{eqnarray}
\label{eq:NLM0}
\Vpix{n} & = & \sum_{r=1}^{R} \abond{r}{n}\Vmat{r} + \bphi_{n}+ \Vnoise_{\nopix}\nonumber\\
 & = &  \MATmat \Vabond{n} + \bphi_{n}+ \Vnoise_{\nopix}, \quad \nopix=1,\ldots,N
\end{eqnarray}
where $\Vmat{\nomat} =
[\mat{\nomat}{1},\ldots,\mat{\nomat}{\nbband}]\transp$ is the
spectrum of the $\nomat$th material present in the scene,
$\abond{r}{n}$ is its corresponding proportion in the $n$th pixel
and $\Vnoise_n$ is an additive independently and non identically distributed
zero-mean Gaussian noise sequence with diagonal covariance matrix
$\bSigma_0=\diag{\bsigma^2}$, denoted as $\Vnoise_n \sim
\Ndistr{\Vzero_{\nbband},\bSigma_0}$, where
$\bsigma^2=[\sigma_1^2,\ldots,\sigma_L^2]\transp$ is the vector of
the $L$  noise variances and $\diag{\bsigma^2}$ is an $L \times L$
diagonal matrix containing the elements of the vector $\bsigma^2$.
Moreover, the term $\bphi_{n}=[\phi_{1,n},\ldots,\phi_{L,n}]\transp$
in \eqref{eq:NLM0} is an unknown $L \times 1$ additive perturbation
vector modeling nonlinear effects occurring in the $n$th pixel. Note
that the usual matrix and vector notations $\MATmat =
[\Vmat{1},\ldots,\Vmat{\nbmat}]$ and
$\Vabond{n}=[\abond{1}{n},\ldots, \abond{\nbmat}{n}]\transp$ have
been used in the second row of Eq. \eqref{eq:NLM0}. There are several motivations for
considering the mixing model \eqref{eq:NLM0}. First, 1) this model reduces to the classical linear mixing model
(LMM) for $\bphi_{n}=\Vzero_{L}$, 2) the model \eqref{eq:NLM0} is general
enough to handle different of kinds of nonlinearities such as the
bilinear model studied in \cite{Fan2009} (referred to as Fan
model (FM)), the generalized bilinear model (GBM) \cite{Halimi2010},
and the polynomial post-nonlinear mixing model (PPNMM) studied for
nonlinear spectral unmixing in \cite{Altmann2012a} and nonlinearity
detection in \cite{Altmann2012b}. These models assume that the mixing
model consists of a linear contribution of the endmembers, corrupted
by at least one additive term characterizing the nonlinear effects.
In the proposed model, all additive terms are gathered in the vector
$\bphi_{n}$. Note that a similar model, called robust LMM, has been
also introduced in \cite{Dobigeon2013whispers}.

Due to physical considerations, the abundance vectors $\Vabond{n}$
satisfy the following positivity and sum-to-one constraints
\begin{equation}
\sum_{\nomat=1}^{\nbmat}{\abond{r}{n}}=1,\ \ \abond{r}{n} >
0, \forall \nomat \in \left\lbrace 1,\ldots,\nbmat \right\rbrace.
\label{eq:abconst}
\end{equation}
The problem addressed in this paper consists of the joint estimation of the abundance vectors
and the detection of nonlinearly mixed pixels (characterized by $\bphi_{n} \neq \Vzero_{L}$).
The two next sections present the proposed Bayesian model for joint unmixing and nonlinearity
detection.

%%%%%%%%%%%%%%%%%%%%%%%%%%%%%%%%%%%%%%%%%%%%%%%%%%%%%%%%%%%%%%%%%%%%%
%%%%%%%%%%%%%%%%%%%%%%%%%%%%%%%%%%%%%%%%%%%%%%%%%%%%%%%%%%%%%%%%%%%%%
\section{Bayesian linear model}

\label{sec:bayesian} The unknown parameter vector associated with
the proposed model \eqref{eq:NLM0} contains the abundances
$\MATabond=[\Vabond{1},\ldots,\Vabond{N}]$ (satisfying the
constraints \eqref{eq:abconst}), the nonlinear terms of each
pixel $\left \lbrace \bphi_{n} \right \rbrace_{n=1,\ldots,N}$, and
the noise variance vector $\bsigma^2$. This section summarizes the
likelihood and the parameter priors associated with the parameters
of the linear part of the model, i.e.,
$\MATabond=[\Vabond{1},\ldots,\Vabond{N}]$ and $\bsigma^2$. One of the main contributions of this paper is the characterization of the nonlinearities that will addressed later in Section
\ref{sec:nonlinear}.

\subsection{Likelihood}
Equation \eqref{eq:NLM0} shows that $\Vpix{n}|\MATmat,\Vabond{n},
\bphi_{n}, \bsigma^2$ is distributed according to a Gaussian
distribution with mean $\MATmat\Vabond{n} + \bphi_n$ and covariance
matrix $\bSigma_0$, denoted as $\Vpix{n}|\MATmat,\Vabond{n},
\bphi_{n}, \bsigma^2 \sim
\Ndistr{\MATmat\Vabond{n}+\bphi_n,\bSigma_0}$. Assuming independence
between the observed pixels, the joint
likelihood of the observation matrix $\MATpix$ can be expressed as\\
$f(\MATpix|\MATmat,\MATabond, \bPhi, \bsigma^2)$
\begin{eqnarray}
\label{eq:likelihood}
     &\propto & |\bSigma_0|^{-N/2}\etr\left[-\dfrac{(\MATpix-\bX)\transp\bSigma_0^{-1}(\MATpix-\bX)}{2}\right]
\end{eqnarray}
where $\bPhi=[\bphi_{1},\ldots,\bphi_{N}]\transp$ is an $L \times N$
nonlinearity matrix, $\propto$ means ``proportional to'',
$\etr(\cdot)$ denotes the exponential trace and $\bX=\MATmat
\MATabond + \bPhi$ is an $L \times N$ matrix.

\subsection{Prior for the abundance matrix $\MATabond$}
Each abundance vector can be written as
$\Vabond{n}=[\Vcoeff{n}\transp,\abond{R}{n}]\transp$
with $\Vcoeff{n}=[\abond{1}{n},\ldots,
\abond{\nbmat-1}{n}]\transp$ and
$\abond{R}{n}=1-\sum_{\nomat=1}^{\nbmat-1}{\abond{r}{n}}$. The LMM
constraints \eqref{eq:abconst} impose that $\Vcoeff{n}$
belongs to the simplex
\begin{equation}
\label{eq:simplex} \Simplex = \left\lbrace \Vcoeff{} \left|
c_{\nomat}> 0,\forall r\in 1,\ldots,R-1,
\sum_{\nomat=1}^{\nbmat-1}{c_{\nomat}} < 1 \right\rbrace
\right.
\end{equation} To reflect the lack of prior knowledge about
the abundances, we propose to assign noninformative prior
distributions to the $N$ vectors $\Vcoeff{n}$. More precisely,
the following uniform prior
\begin{eqnarray}
\label{eq:prior_abond} f(\Vcoeff{n}) \propto
\Indicfun{\Simplex}{\Vcoeff{n}},\quad n \in \left \lbrace
1,\ldots,N\right \rbrace
\end{eqnarray}
is assigned to each vector $\Vcoeff{n}$, where $\Indicfun{\Simplex}{\cdot}$ is the indicator function defined on the simplex $\Simplex$. Assuming prior independence
between the $N$ abundance vectors $\left \lbrace \Vabond{n} \right
\rbrace_{n=1,\ldots,N}$ leads to the following joint prior
distribution
\begin{eqnarray}
\label{eq:joint_prior_coeff} f(\MATcoeff)  =
\prod_{n=1}^{N}f(\Vcoeff{n})
\end{eqnarray}
where $\MATcoeff=[\Vcoeff{1},\ldots,\Vcoeff{N}]$ is an $(R-1)
\times N$ matrix.

\subsection{Prior for the noise variance vector $\bsigma^2$}
A Jeffreys' prior is chosen for the noise variance of each spectral
band $\sigma_{\ell}^2$
\begin{eqnarray}
    \label{eq:priorvar}
    f(\sigma_{\ell}^2) \propto \dfrac{1}{\sigma_{\ell}^2} \Indicfun{\R^+}{\sigma_{\ell}^2}
\end{eqnarray}
which reflects the absence of knowledge for this parameter (see
\cite{Bernardo94} for motivation). Assuming prior independence
between the noise variances, we obtain
\begin{eqnarray}
    \label{eq:jointpriorvar}
    f(\bsigma^2) = \prod_{\ell=1}^L f(\sigma_{\ell}^2).
\end{eqnarray}

%%%%%%%%%%%%%%%%%%%%%%%%%%%%%%%%%%%%%%%%%%%%%%%%%%%%%%%%%%%%%%%%%%%%%
%%%%%%%%%%%%%%%%%%%%%%%%%%%%%%%%%%%%%%%%%%%%%%%%%%%%%%%%%%%%%%%%%%%%%

\section{Modeling the nonlinearities}

\label{sec:nonlinear}

%\subsection{Introducing spatial dependencies between nonlinearity vectors}
We propose in this paper to exploit spatial correlations between the
pixels of the hyperspectral image to be analyzed. It seems
reasonable to assume that nonlinear effects occurring in a given
pixel are related to the nonlinear effects present in
neighboring pixels. Formally, the hyperspectral image is assumed to
be partitioned into $K$ classes denoted as $\calC_0,\ldots,\calC_{K-1}$.
Let $\mathcal{I}_k \subset {1, \ldots , N}$ denote the subset of
pixel indexes belonging to the $k$th class ($k=0,\ldots,K-1$). An $N
\times 1$ label vector $\bz = [z_1, \ldots,z_N ]\transp$ with $z_n
\in \{0,\ldots, K-1\}$ is introduced to identify the class of each
image pixel, i.e.,
\begin{equation}\Vpix{n} \in C_k \Leftrightarrow
n \in \mathcal{I}_k \Leftrightarrow z_n=k.
\end{equation}
In each class, nonlinearity vectors to be estimated are assumed to
share the same statistical properties, as will be shown in the sequel.

\subsection{Prior distribution for the nonlinearity matrix $\bPhi$}

As mentioned above, the mixing model \eqref{eq:NLM0} reduces to the
LMM for $\bphi_{n}=\Vzero_{L}$. For nonlinearity detection, it makes
sense to consider a pixel class (referred to as class $\calC_0$)
corresponding to linearly mixed pixels. The resulting prior
distribution for $\bphi_{n}$ conditioned upon $z_n=0$ is given by
\begin{eqnarray}
f(\bphi_{n} | z_n=0) = \prod_{\ell=1}^{L}\delta(\phi_{\ell,n}).
\end{eqnarray}
It can be seen that bilinear models and more generally polynomial
models (i.e., model involving polynomials nonlinearities with
respect to the endmembers) are particularly well adapted to model
scattering effects, mainly observed in vegetation and urban areas.
Consequently, it makes sense to assume that the nonlinearities
$\bphi_{n}$ depend on the endmember matrix $\MATmat$. Nonlinear
effects can vary, depending on the relief of the scene, the
underlying components involved in the mixtures and the observation
conditions to name a few factors. This makes the choice of a single
informative prior distribution challenging. From a classification
point of view, it is interesting to identify regions or classes
where similar nonlinearities occur. For these reasons, we propose to
divide nonlinearly mixed pixels into $K-1$ classes and to assign
different priors for the nonlinearity vectors belonging to the different
classes. The nonlinearities (of nonlinearly mixed
pixels) are assumed to be random. Assume $\Vpix{n}$ belongs to the
$k$th class. The prior distribution of the corresponding nonlinear
term $\bphi_{n}$ is given by the following GP ($k=1,\ldots, K-1$)
\begin{eqnarray}
\label{eq:mu_prior0} \bphi_{n}|\MATmat,z_n=k,s_k^2 \sim
\Ndistr{\Vzero_{\nbband},s_k^2\bK_{\MATmat}},
\end{eqnarray}
where $\bK_{\MATmat}$ is an $L \times L$ covariance matrix
parameterized by the endmember matrix $\MATmat$ and $s_k^2$ is a
scaling hyperparameter that tunes the energy of the nonlinearities
in the $k$th class. Note that all nonlinearity vectors within the
same class share the same prior. The performance of the unmixing
procedure depends on the choice of $\bK_{\MATmat}$, more precisely
on the similarity measure associated with the covariance matrix. In
this paper, we consider the symmetric second order polynomial kernel, which has received considerable interest in the machine learning community \cite{Scholkopf2001}. This kernel is defined as follows
\begin{eqnarray}
%\label{eq:bK} \bK_{\MATmat}=\left(\MATmat
%\MATmat\transp\right)\odot\left(\MATmat \MATmat\transp\right)
\label{eq:bK}
\left[\bK_{\MATmat} \right]_{i,j} = \left(\Vmat{i,:}\transp\Vmat{j,:} \right)^2, \quad i,j \in \left\lbrace 1,\ldots,\nbband \right\rbrace,
\end{eqnarray}
where $\odot$ denotes the Hadamard (termwise) product and $\Vmat{i,:}$ denotes the $i$th row of $\MATmat$.  Polynomial kernels are particularly well adapted to characterize multiple scattering effects (modeled by polynomial functions of the endmembers). Note that the parametrization of the matrix $\bK_{\MATmat}$ in \eqref{eq:bK} only involves bilinear and quadratic terms\footnote{Note: it can be shown that \eqref{eq:mu_prior0} and \eqref{eq:bK} can be obtained by defining $\bphi_{n}$ as a linear combination of terms $\Vmat{i}\odot\Vmat{j}$ (as in \cite{Halimi2010}) and by marginalizing the corresponding coefficients using a Gaussian prior parameterized by $s_k^2$. Marginalizing these coefficients allows the number of unknown parameters to be significantly reduced, leading to the nonlinearities being characterized by a single parameter $s_k^2$.}
with respect to the endmembers $\Vmat{r},r=1,\ldots,R$. More, precisely, the matrix $\bK_{\MATmat}$ can be rewritten as
\begin{equation*}
  \bK_{\MATmat}=\bQ \bQ\transp
\end{equation*}
where
$\bQ=[\Vmat{1}\odot\Vmat{1},\ldots,\Vmat{R}\odot\Vmat{R},
\sqrt{2}\Vmat{1}\odot\Vmat{2},\ldots,\sqrt{2}\Vmat{R-1}
\odot\Vmat{R}]$
is an $L \times R(R+1)/2$ matrix. Note also that a polynomial kernel similar to \eqref{eq:bK} has been recently considered in \cite{Chen2012} and that other kernels such as the Gaussian kernel could be investigated to model other nonlinearities as in \cite{Kalaitzis2012}.

\subsection{Prior distribution for the label vector $\bz$}

%\subsubsection{Markov Random Field}

\begin{figure}[h!]
  \centering
  \includegraphics[width=\columnwidth]{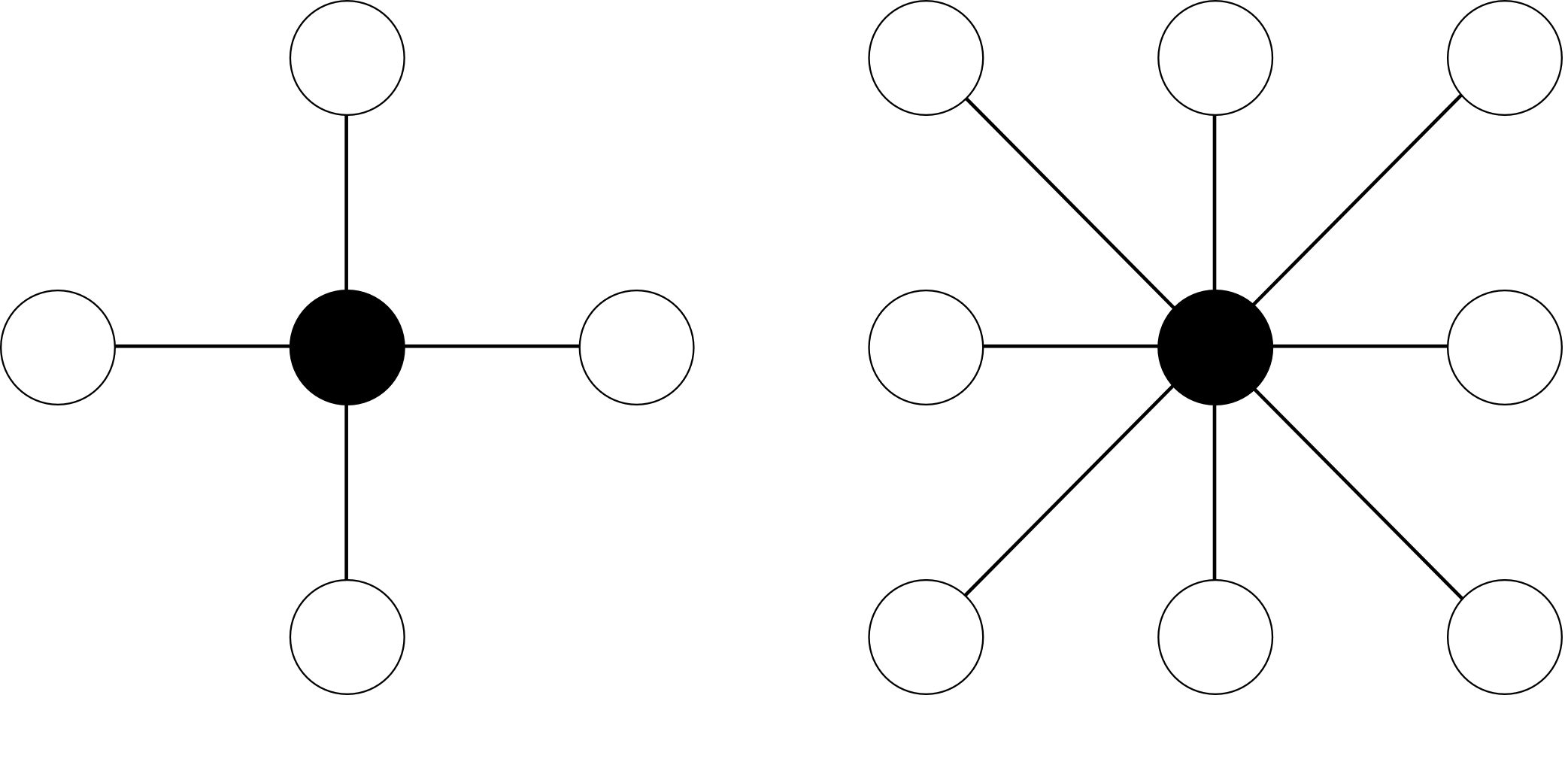}
  \caption{4-pixel (left) and 8-pixel (right) neighborhood structures. The
considered pixel appear as a black circle whereas its neighbors are
depicted in white.}
  \label{fig:neighbor_struct}
\end{figure}

In the context of hyperspectral image analysis, the labels
$z_1,\ldots,z_N$ indicate the pixel classes and take values in
$\{0,\ldots,K-1\}$ where $K$ is the number of classes and the set $\{
z_n\}_{n=1,\ldots,N}$ forms a random field. To exploit the
correlation between pixels, a Markov random field is introduced as a
prior distribution for $z_n$ given its neighbors
$\bz_{\mathcal{V}(n)}$ , i.e.,
\begin{equation}
f(z_n|\bz_{\backslash n}) = f (z_n|\bz_{\mathcal{V}(n)})
\end{equation}
where $\mathcal{V}(n)$ is the neighborhood of the $n$th pixel and
$\bz_{\backslash n} = \{z_{n'}\}_{n' \neq n}$. More precisely, this
paper focuses on the Potts-Markov model since it is very appropriate
for hyperspectral image segmentation \cite{Eches2011}. Given a
discrete random field $\bz$ attached to an image with $N$ pixels,
the Hammersley-Clifford theorem yields
\begin{eqnarray}
\label{eq:MRF_prior} f (\bz) = \dfrac{1}{G(\beta)} \exp \left[\beta
\sum_{n=1}^{N} \sum_{n' \in \mathcal{V}(n)} \delta(z_n -
z_{n'})\right]
\end{eqnarray}
where $\beta>0$ is the granularity coefficient, $G(\beta)$ is a
normalizing (or partition) constant and $\delta(\cdot)$ is the Dirac
delta function. Several neighborhood structures can be employed to
define $\mathcal{V}(n)$. Fig. \ref{fig:neighbor_struct} shows two
examples of neighborhood structures. The eight pixel structure (or
2-order neighborhood) will be considered in the rest of the paper.

The hyperparameter $\beta$ tunes the degree of homogeneity of each
region in the image. More precisely, small values of $\beta$ yield
an image with a large number of regions, whereas large values of
$\beta$ lead to fewer and larger homogeneous regions. In this paper,
the granularity coefficient is assumed to be known. Note however
that it could be also included within the Bayesian model and estimated
using the strategy described in \cite{Pereyra2013ip}.

%Once the neighborhood structure has been set, the Markov random
%field (MRF) can be introduced.
% The proposed MRF will
%be presented in detail in Section \ref{sec:bayesian}. The next
%section presents the Bayesian model associated with the nonlinear
%model \eqref{eq:NLM0} for nonlinear spectral unmixing and
%nonlinearity detection.

%\subsubsection{Label vector $\bz$}

\subsection{Hyperparameter priors}
The performance of the proposed Bayesian model for spectral unmixing
mainly depends on the values of the hyperparameters $\left \lbrace
s_k^2 \right \rbrace_{k=1,\ldots,K}$. When the hyperparameters are
difficult to adjust, it is the norm to include them in the unknown
parameter vector, resulting in a hierarchical Bayesian model
\cite{Dobigeon2009,Altmann2012a}. This strategy requires the
definition of prior distributions for the hyperparameters.

The following inverse-gamma prior distribution
\begin{eqnarray}
s_k^2|\gamma,\nu \sim \mathcal{IG}(\gamma,\nu), \quad \forall k \in
\{1,\ldots,K\}
\end{eqnarray}
is assigned to the nonlinearity hyperparameters, where
$(\gamma,\nu)$ are additional parameters that will be fixed to ensure a noninformative prior for $s_k^2$ ($(\gamma,\nu)=(1,1/4)$ in all simulations presented in this paper).
Assuming prior independence between the hyperparameters, we obtain
\begin{eqnarray}
    \label{eq:jointpriorhyperparam}
    f(\bs^2|\gamma,\nu) = \prod_{k=1}^{K-1} f(s_k^2|\gamma,\nu).
\end{eqnarray}
where $\bs^2=[s_1^2,\ldots,s_K^2]\transp$.

%%%%%%%%%%%%%%%%%%%%%%%%%%%%%%%%%%%%%%%%%%%%%%%%%%%%%%%%%%%%%%%%%%%%%
%%%%%%%%%%%%%%%%%%%%%%%%%%%%%%%%%%%%%%%%%%%%%%%%%%%%%%%%%%%%%%%%%%%%%
\section{Bayesian inference using a Metropolis-within-Gibbs sampler}
\label{sec:Gibbs}

\subsection{Marginalized joint posterior distribution}
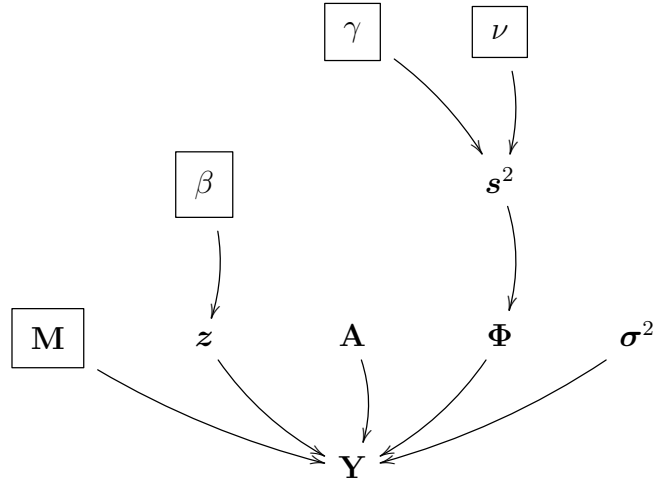
\begin{figure}[h!]
\centerline{ \xymatrix{
 & & *+<0.05in>+[F-]+{\gamma} \ar@/^/[rd]& *+<0.05in>+[F-]+{\nu} \ar@/^/[d]&  \\
  & *+<0.05in>+[F-]+{\beta} \ar@/^/[d] &  & \bs^2 \ar@/^/[d] & \\
  *+<0.05in>+[F-]+{\MATmat} \ar@/_/[rrd]    & \bz \ar@/_/[rd] &  \MATabond  \ar@/^/[d] &  \bPhi \ar@/^/[ld]& \bsigma^2 \ar@/^/[lld]  \\
  & & \MATpix &   & }
} \caption{DAG for the parameter and hyperparameter priors (the
fixed parameters appear in boxes).} \label{fig:DAG}
\end{figure}

The resulting directed acyclic graph (DAG) associated with the
proposed Bayesian model introduced in Sections \ref{sec:bayesian}
and \ref{sec:nonlinear} is depicted in Fig. \ref{fig:DAG}.

Assuming prior independence between $\MATabond,(\bPhi,\bz)$ and
$\bsigma^2$, the posterior distribution of $(\bPhi,\paramvect)$ where
$\paramvect=(\MATcoeff,\bz,\bsigma^2,\bs^2)$ can be expressed as
\begin{eqnarray}
f(\paramvect,\bPhi|\MATpix,\MATmat) & \propto& f(\MATpix|\MATmat,\paramvect,\bPhi)f(\bPhi|\MATmat,\bz,\bs^2)f(\paramvect),\nonumber
\end{eqnarray}
where $f(\paramvect)=f(\MATcoeff)f(\bsigma^2)f(\bz)f(\bs^2)$.
This distribution can be marginalized with respect to $\bPhi$ as follows
\begin{eqnarray}
\label{eq:posterior}
f(\paramvect|\MATpix,\MATmat) & \propto&  f(\paramvect) \int f(\MATpix|\MATmat,\paramvect,\bPhi)f(\bPhi|\MATmat,\bz,\bs^2) \mathrm{d} \bPhi\nonumber\\
& \propto & f(\paramvect) f(\MATpix|\MATmat,\paramvect)
\end{eqnarray}
where
\begin{eqnarray}
 f(\MATpix|\MATmat,\paramvect)&  = & \int f(\MATpix|\MATmat,\paramvect,\bPhi)f(\bPhi|\MATmat,\bz,\bs^2) \mathrm{d} \bPhi\\
 & \propto & \prod_{k=0}^{K-1}\prod_{n \in \mathcal{I}_k} \dfrac{1}{ |\bSigma_k|^{\frac{1}{2}}}
\exp \left[-\dfrac{1}{2}
\bar{\Vpix{}}_{n}\transp\bSigma_k^{-1}\bar{\Vpix{}}_{n}\right]\nonumber
\end{eqnarray}
with $\bSigma_0=\diag{\bsigma^2}$, $\bSigma_k=s_k^2\bK_{\MATmat}
+ \bSigma_0$ ($k=1,\ldots,K-1$)
and $\bar{\Vpix{}}_{n}=\Vpix{n}-\MATmat \Vabond{n}$. The advantage of this marginalization is to avoid sampling the nonlinearity matrix $\bPhi$. Thus, the nonlinearities are fully characterized by the known endmember matrix, the class labels and the values of the hyperparameters in $\bs^2=[s_1^2,\ldots,s_K^2]\transp$.

Unfortunately, it is difficult to obtain closed form expressions for
standard Bayesian estimators associated with \eqref{eq:posterior}.
In this paper, we propose to use efficient Markov Chain Monte Carlo
(MCMC) methods to generate samples asymptotically distributed
according to \eqref{eq:posterior}. The next part of this section
presents the Gibbs sampler which is proposed to sample according to
\eqref{eq:posterior}. The principle of the Gibbs sampler is to
sample according to the conditional distributions of the posterior
of interest \cite[Chap. 10]{Robert2004}. Due to the large number of
parameters to be estimated, it makes sense to use a block Gibbs
sampler to improve the convergence of the sampling procedure. More
precisely, we propose to sample sequentially the $N$ labels in
$\bz$, the abundance matrix $\MATabond$, the noise variances
$\bsigma^2$ and $\bs^2$ using moves that are detailed in the next
paragraphs.

\subsection{Sampling the labels}
For the $n$th pixel ($n \in \{1,\ldots,N\}$), the label $z_n$ is a
discrete random variable whose conditional distribution is fully
characterized by the probabilities
\begin{eqnarray}
P(z_n=k|\Vpix{n},\MATmat,\paramvect_{\backslash z_n}) & \propto&  f(\Vpix{n}|\MATmat,\bs^2,z_n=k,\Vabond{n})\nonumber\\
 & \times & f(z_n|\bz_{\backslash n}),\nonumber
\end{eqnarray}
where $\paramvect_{\backslash z_n}$ denotes $\paramvect$ without
$z_n$, $k = 0, \ldots,K-1$ (for K classes). These posterior
probabilities are
$$P(z_n=k|\Vpix{n},\MATmat,\paramvect_{\backslash z_n}) \propto \exp \left[\beta \sum_{p=1}^{N} \sum_{p' \in \mathcal{V}(p)} \delta(z_p - z_{p'})\right]$$
\begin{eqnarray}
\label{eq:post_z}
\times \dfrac{1}{ |\bSigma_k|^{\frac{1}{2}}}
\exp \left[-\dfrac{1}{2} \bar{\Vpix{}}_{n}\transp\bSigma_k^{-1}\bar{\Vpix{}}_{n}\right].
\end{eqnarray}
Consequently, sampling $z_n$ from its conditional distribution can
be achieved by drawing a discrete value in the finite set
$\{0,\ldots,K-1\}$ with the probabilities defined in
\eqref{eq:post_z}.

\subsection{Sampling the abundance matrix $\MATabond$}
 \label{subsec:sample_z}
Sampling from $f(\MATcoeff|\MATpix,\MATmat,\bz,\bsigma^2,\bs^2)$
seems difficult due to the complexity of this distribution. However,
it can be shown that
\begin{eqnarray}
f(\MATcoeff|\MATpix,\MATmat,\bz,\bsigma^2,\bs^2) = \prod_{n=1}^{N} f(\Vcoeff{n}|\Vpix{n},\MATmat,z_n,\bsigma^2,\bs^2),
\end{eqnarray}
i.e., the $N$ abundance vectors $\left \lbrace \Vabond{n} \right
\rbrace_{n=1,\ldots,N}$ are a posteriori independent and can be
sampled independently in a parallel manner. Straightforward
computations lead to
\begin{eqnarray}
\label{eq:post_abond}
\Vcoeff{n}|\Vpix{n},\MATmat,z_n=k,\bsigma^2,\bs^2 \sim \mathcal{N}_{\Simplex}(\bar{\bc}_n,\bPsi_n)
\end{eqnarray}
where
\begin{eqnarray}
\bPsi_n & = & \left(\widetilde{\MATmat}\transp\bSigma_k^{-1}\widetilde{\MATmat}\right)^{-1}\nonumber\\
\bar{\bc}_n & = & \bPsi_n \widetilde{\MATmat}\transp\bSigma_k^{-1}\tilde{\Vpix{}}_n\nonumber\\
\widetilde{\MATmat} & = & [\Vmat{1}-\Vmat{R},\ldots,\Vmat{\nbmat-1}-\Vmat{R}]
\end{eqnarray}
and $\tilde{\Vpix{}}_n=\Vpix{n}-\Vmat{R}$. Moreover,
$\mathcal{N}_{\Simplex}(\bar{\bc}_n,\bPsi_n)$ denotes the truncated multivariate Gaussian distribution defined on the simplex $\Simplex$ with hidden mean $\bar{\bc}_n$ and hidden covariance matrix
$\bPsi_n$. Sampling from \eqref{eq:post_abond} can be
achieved efficiently using the method recently proposed in
\cite{Pakman2012}.

\subsection{Sampling the noise variance $\sigma^2$}
It can be shown from \eqref{eq:posterior} that
\begin{eqnarray}
f(\bsigma^2|\MATpix,\MATmat,\MATabond,\bz,\bs^2) = \prod_{\ell=1}^{L} f(\sigma_{\ell}^2|\MATpix,\MATmat,\MATabond,\bz,\bs^2),
\end{eqnarray}
where\\ $f(\sigma_{\ell}^2|\MATpix,\MATmat,\MATabond,\bz,\bs^2)$
\begin{eqnarray}
\label{eq:post_sigma2}
\propto \dfrac{1}{\sigma_{\ell}^2}\prod_{k=0}^{K-1}\prod_{n \in \mathcal{I}_k} \dfrac{1}{ |\bSigma_k|^{\frac{1}{2}}}
\exp \left[-\dfrac{1}{2} \bar{\Vpix{}}_{n}\transp\bSigma_k^{-1}\bar{\Vpix{}}_{n}\right]  \Indicfun{\R^+}{\sigma_{\ell}^2}.
\end{eqnarray}
Sampling from \eqref{eq:post_sigma2} is not straightforward. In this case,
an accept/reject procedure can be used to update $\sigma_{\ell}^2$, leading to a hybrid Metropolis-within-Gibbs sampler. In this paper, we introduce
the standard change of variable
$\delta_{\ell}=\log(\sigma_{\ell}^2)$, $\delta_{\ell} \in \R$. A Gaussian random walk for $\delta_{\ell}$ is used to update
the variance $\sigma_{\ell}^2$. Note that the noise variances are a
posteriori independent. Thus they can be updated in a parallel
manner. The variances of the $L$ parallel Gaussian random walk
procedures have been adjusted during the burn-in period of the
sampler to obtain an acceptance rate close to $0.5$, as recommended
in \cite[p. 8]{Robertmcmc}.

\subsection{Sampling the vector $\bs^2$}
It can be shown from \eqref{eq:posterior} that
\begin{eqnarray}
f(\bs^2|\MATpix,\MATmat,\MATabond,\bz,\bsigma^2,\gamma,\nu) = \prod_{k=1}^{K-1} f(s_k^2|\MATpix,\MATmat,\MATabond,\bsigma^2,\gamma,\nu),\nonumber
\end{eqnarray}
where\\ $f(s_k^2|\MATpix,\MATmat,\MATabond,\bsigma^2,\gamma,\nu)$
\begin{eqnarray}
\label{eq:post_deltan}
 \propto f(s_k^2|\gamma,\nu) \prod_{n \in \mathcal{I}_k} \dfrac{1}{ |\bSigma_k|^{\frac{1}{2}}}
\exp \left[-\dfrac{1}{2} \bar{\Vpix{}}_{n}\transp\bSigma_k^{-1}\bar{\Vpix{}}_{n}\right].
\end{eqnarray}
Due to the complexity of the conditional distribution
\eqref{eq:post_deltan}, Gaussian random walk procedures are used in
the log-space to update the hyperparameters
$\{s_k^2\}_{k=1,\ldots,K-1}$ in a parallel manner (similarly to the
noise variance updates). Again, the proposal variances are adjusted
during the burn-in period of the sampler.

After generating $N_{\textrm{MC}}$ samples using the procedures
detailed above and removing $N_{\textrm{bi}}$ iterations associated
with the burn-in period of the sampler ($N_{\textrm{bi}}$ has been
set from preliminary runs), the marginal maximum a posteriori (MAP)
estimator of the label vector, denoted as
$\hat{\bz}_{\textrm{MAP}}$, can be computed. The label vector
estimator is then used to compute the minimum mean square error
(MMSE) of $\MATabond$ conditioned upon
$\bz=\hat{\bz}_{\textrm{MAP}}$. Finally, the noise variances and the
hyperparameters $\{s_k^2\}_{k=1,\ldots,K-1}$ are estimated using the empirical averages of the generated samples (MMSE
estimates). The next section studies the performance of the proposed
algorithm for synthetic hyperspectral images.

\section{Simulations for Synthetic data}
\label{sec:simu_synth}
\subsection{First scenario: RCA vs. linear unmixing}
\begin{figure}[h!]
  \centering
  \includegraphics[width=\columnwidth]{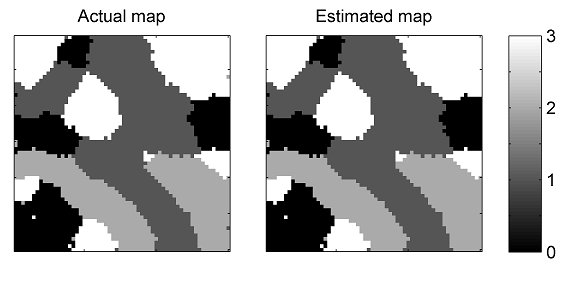}
  \caption{Actual (left) and estimated (right) classification maps of the synthetic image associated with the first scenario.}
  \label{fig:detection_RCA_synth}
\end{figure}

\begin{figure}[h!]
  \centering
  \includegraphics[width=\columnwidth]{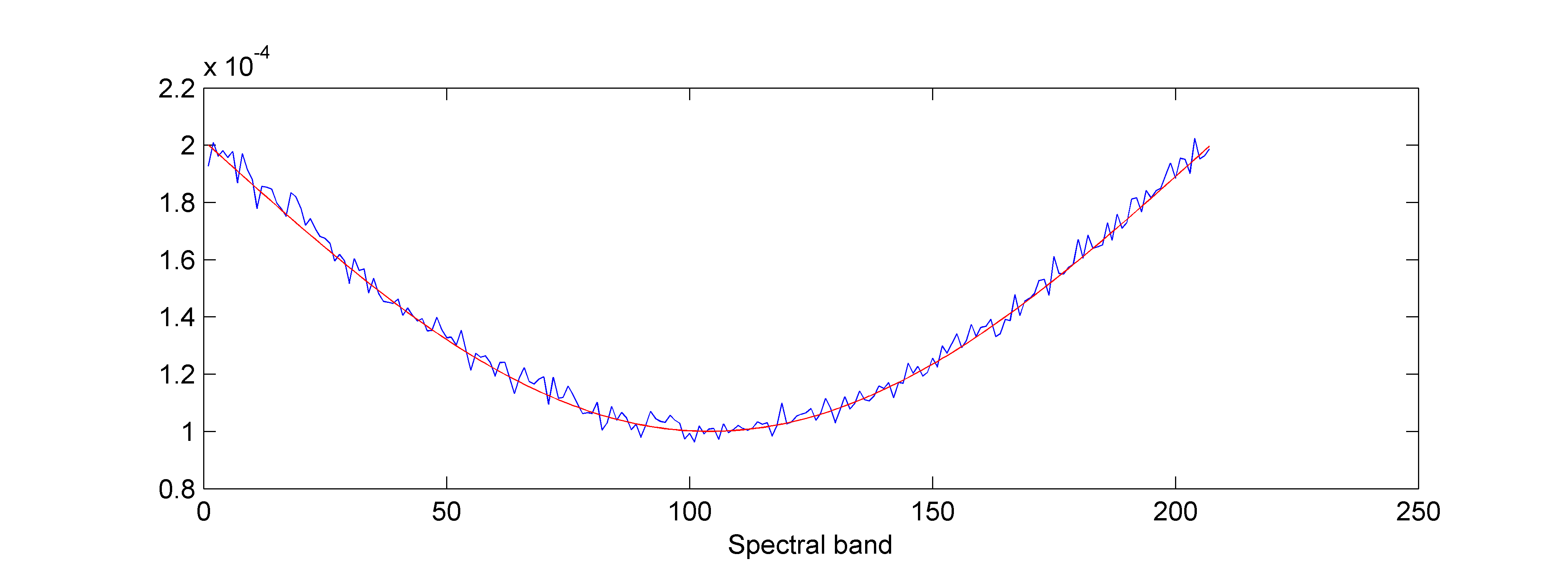}
  \caption{Actual noise variances (red) and variances estimated by the RCA-SU algorithm (blue) for the synthetic image associated with the first scenario.}
  \label{fig:Noise_variances_RCA_synth}
\end{figure}

The performance of the proposed joint nonlinear SU and nonlinearity
detection algorithm is first evaluated by unmixing a synthetic image
of $60 \times 60$ pixels generated according to the model
\eqref{eq:NLM0}. The $\nbmat = 3$ endmembers contained in these images (i.e., green grass, olive green paint
and galvanized steel metal) have $L=207$
different spectral bands and have been
extracted from the spectral libraries provided with the ENVI
software \cite{ENVImanual2003} . The number of classes has been set to
$K=4$, i.e, $K-1=3$ classes of nonlinearly mixed pixels. The
hyperparameters $\left \lbrace
s_k^2 \right \rbrace_{k=1,\ldots,3}$ have been fixed as shown in Table
\ref{tab:class_variances}, which represents three possible levels of
nonlinearity. For each class, the nonlinear terms have been generated
according to \eqref{eq:mu_prior0}. The label map generated
with $\beta = 1.2$ is shown in Fig. \ref{fig:detection_RCA_synth}
(left). The abundance vectors $\Vabond{n}, n=1,\ldots,3600$ have
been randomly generated according to a uniform distribution over the
admissible set defined by the positivity and sum-to-one constraints. The noise variance (depicted in Fig.
\ref{fig:Noise_variances_RCA_synth} as a function of the spectral
bands) have been arbitrarily fixed using
\begin{eqnarray}
\sigma_{\ell}^2= 10^{-4} \left[2-\sin \left( \pi \dfrac{\ell}{L-1}\right)\right].
\end{eqnarray}
to model a non-i.i.d. (colored) noise. The joint nonlinear SU
and nonlinearity detection algorithm, denoted as ``RCA-SU'', has been applied to this data set with
$N_{\textrm{MC}}=3000$ and $N_{\textrm{bi}}=1000$. Fig.
\ref{fig:detection_RCA_synth} (right) shows that the estimated label map
(marginal MAP estimates) is in agreement with the actual label
map. Moreover, the confusion matrix depicted in Table \ref{tab:confusion_matrix} illustrate the performance of the RCA-SU in term of pixel classification. Table \ref{tab:class_variances} shows that the RCA-SU
provides accurate hyperparameter estimates and thus can be used to obtain information about the importance of
nonlinearities in the different regions. Note that the estimation error is
computed using $|s_k^2-\hat{s}_k^2|/s_k^2$, where $s_k^2$
and $\hat{s}_k^2$ are the actual and estimated dispersion parameters for the $k$th class. The estimated noise
variances, depicted in Fig. \ref{fig:Noise_variances_RCA_synth} are also in
good agreement with the actual values of the variances.
The quality  of abundance estimation can be evaluated by comparing
the estimated and actual abundance vectors using the root normalized
mean square error (RNMSE) defined in each class by
\begin{eqnarray}
\label{eq:RMSE}
    \textrm{RNMSE}_k= \sqrt{\dfrac{1}{N_k \nbmat}\sum_{n \in \mathcal{I}_k}
    {\norm{\hat{\Vabonds}_{\nopix} - \Vabond{\nopix}}^2}}
\end{eqnarray}
with  $N_k=\textrm{card}(\mathcal{I}_k)$ and where $\Vabond{\nopix}$
and $\hat{\Vabonds}_{\nopix}$ are the actual and estimated abundance
vectors for the $\nopix$th pixel of the image. For
this scenario, the proposed algorithm is compared with the classical
FCLS algorithm \cite{Heinz2001} assuming the LMM. Comparisons to
nonlinear SU methods will be addressed in the next paragraph
(scenario 2). Table
\ref{tab:RNMSE_synth1} shows the RNMSEs obtained with the
proposed and the FLCS algorithms for this first data set. These
results show that the two algorithms provide similar abundance
estimates for the first class, corresponding to linearly mixed pixels. For
the three nonlinear classes, the estimation performance is reduced. However,  the proposed algorithm provides better results than the
FCLS algorithm that does not handle nonlinear effects.
\begin{table}[h!]
\renewcommand{\arraystretch}{1.2}
\begin{footnotesize}
\begin{center}
\caption{First scenario: Confusion matrix ($N=3600$ pixels).\label{tab:confusion_matrix}}
\begin{tabular}{|c|c|c|c|c|c|}
\cline{3-6}
\multicolumn{2}{c|}{} &  \multicolumn{4}{c|}{Estimated classes}  \\
\cline{3-6}
\multicolumn{2}{c|}{} & $\calC_0$ & $\calC_1$ & $\calC_2$ &$\calC_3$ \\
\hline
\multirow{4}*{Actual classes} & $\calC_0$ & $659$ & $0$ & $0$ & $0$\\
\cline{2-6}
 & $\calC_1$ & $1$ & $1274$ & $2$ & $0$\\
\cline{2-6}
 & $\calC_2$ & $0$ & $4$ & $787$ & $2$\\
\cline{2-6}
 & $\calC_3$ &$0$ & $0$&$0$ &$871$ \\
\hline
\end{tabular}
\end{center}
\end{footnotesize}
\vspace{-0.4cm}
\end{table}

\begin{table}[h!]
\renewcommand{\arraystretch}{1.2}
\begin{footnotesize}
\begin{center}
\caption{First scenario: Hyperparameter
estimation.\label{tab:class_variances}}
\begin{tabular}{|c|c|c|c|}
\cline{2-4}
\multicolumn{1}{c|}{} &  $s_1^2$  & $s_2^2$ & $s_3^2$  \\
\hline
Actual value & $0.01$ & $0.1$ & $1$ \\
\hline
Estimation error & $2.76\%$ & $1.12\%$ & $0.28\%$\\
\hline
\end{tabular}
\end{center}
\end{footnotesize}
\vspace{-0.4cm}
\end{table}

\begin{table}[h!]
\renewcommand{\arraystretch}{1.2}
\begin{footnotesize}
\begin{center}
\caption{RNMSEs ($\times 10^{-2}$): synthetic images
.\label{tab:RNMSE_synth1}}
\begin{tabular}{|c|c|c|c|c|}
\cline{2-5}
\multicolumn{1}{c|}{} &  Class \#0  & Class \#1 & Class \#2 & Class \#3  \\
\hline
FCLS & $0.38$ & $15.23$ & $29.95$ & $42.79$\\
\hline
RCA-SU & $\textbf{\blue{0.38}}$ & $\textbf{\blue{2.83}}$ & $\textbf{\blue{3.99}}$ & $\textbf{\blue{4.23}}$ \\
\hline
\end{tabular}
\end{center}
\end{footnotesize}
\vspace{-0.4cm}
\end{table}

\subsection{Second scenario: RCA vs. nonlinear unmixing}
\subsubsection{Data set}
\begin{figure}[h!]
\begin{minipage}[b]{.48\linewidth}
  \centering
% \centerline{\epsfig{figure=image3.ps,width=4.0cm}}
\includegraphics[width=4.8cm]{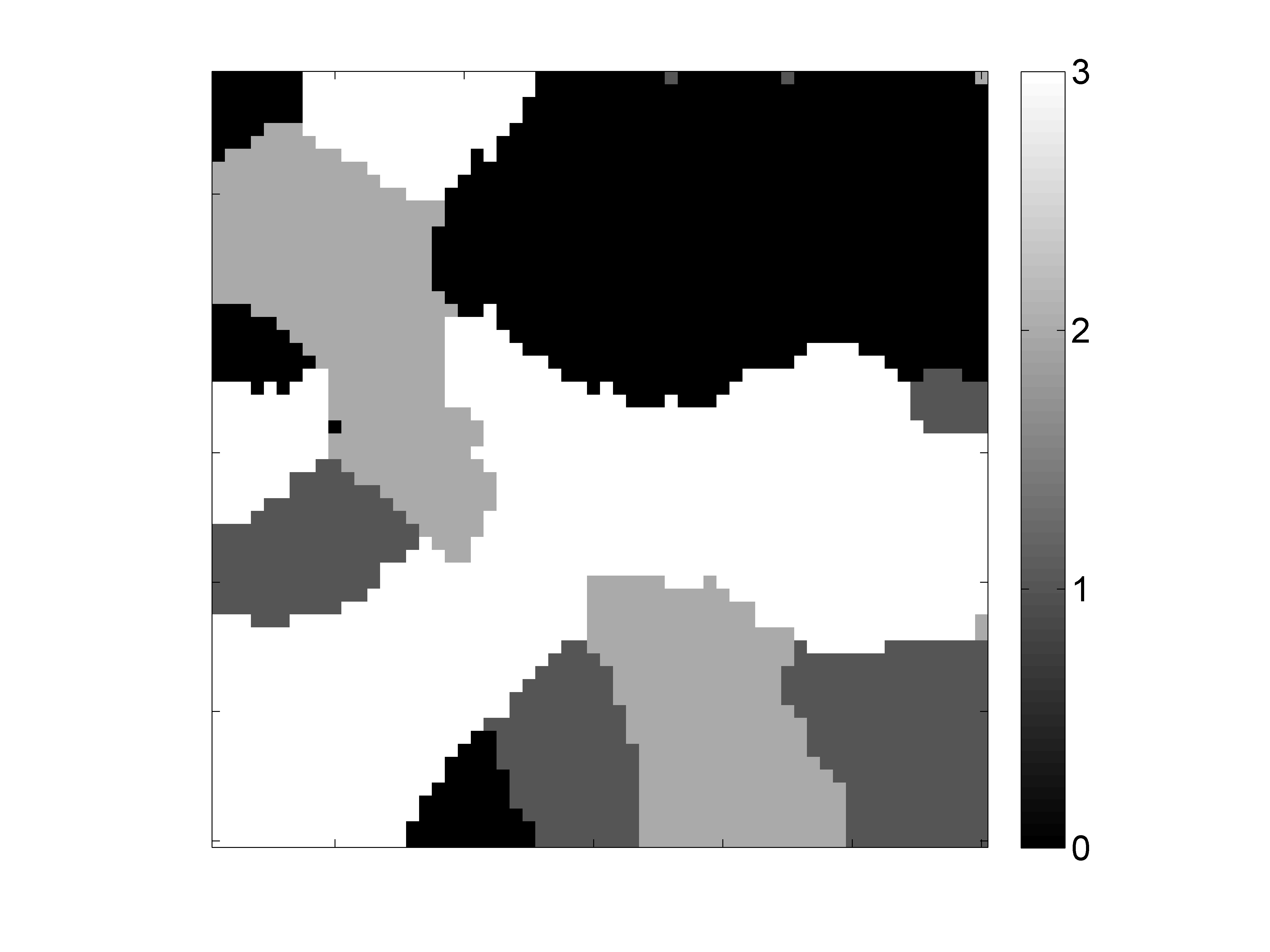}
 % \vspace{-0.2cm}
  \centerline{(a) Actual label map.}\medskip
\end{minipage}
\hfill
\begin{minipage}[b]{0.48\linewidth}
  \centering
% \centerline{\epsfig{figure=image4.ps,width=4.0cm}}
\includegraphics[width=4.8cm]{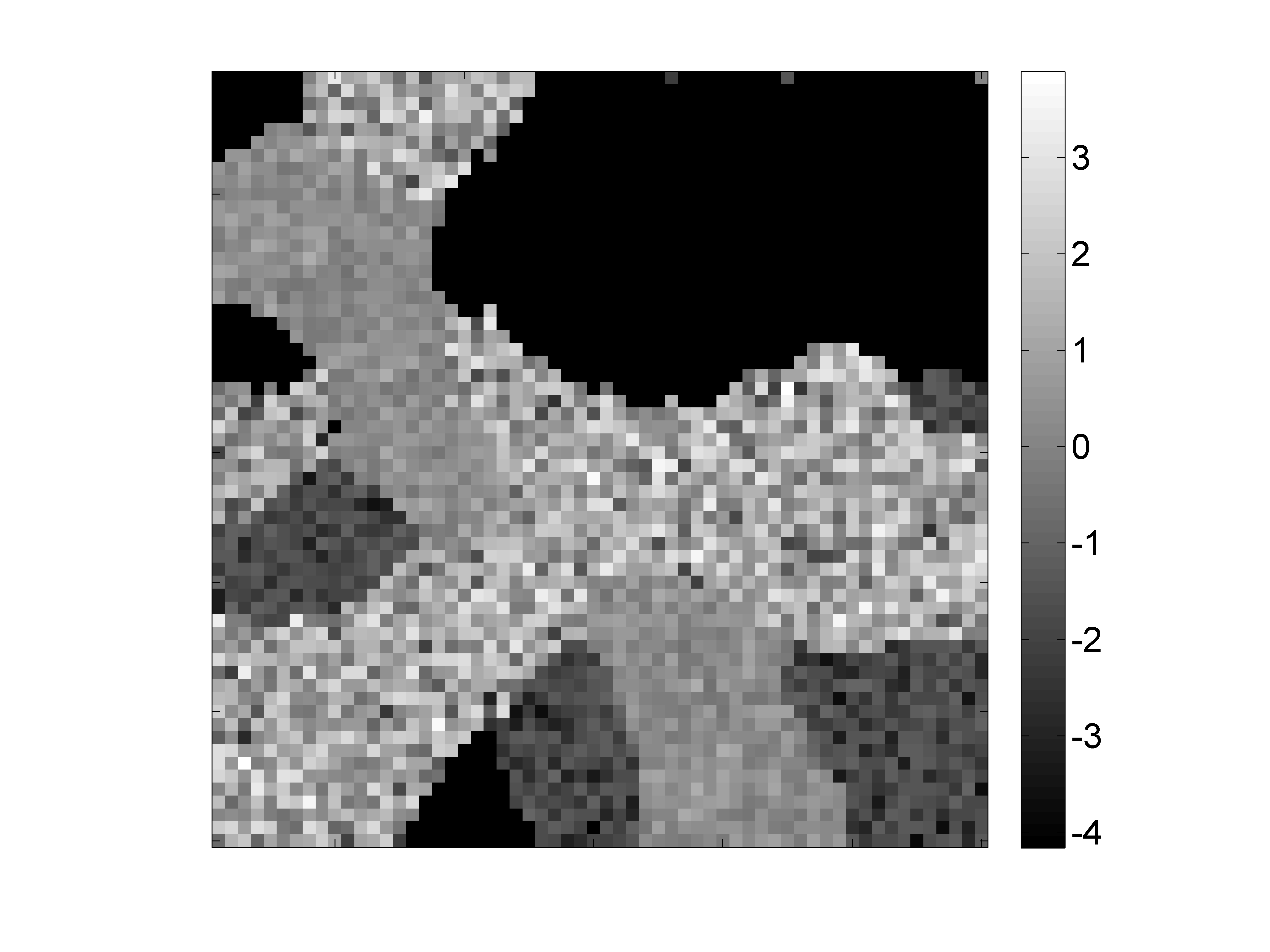}
 % \vspace{-0.2cm}
  \centerline{(b) $\log\left(\norm{\bphi_n}^2\right)$.}\medskip
\end{minipage}
\begin{minipage}[b]{.48\linewidth}
  \centering
% \centerline{\epsfig{figure=image3.ps,width=4.0cm}}
\includegraphics[width=4.8cm]{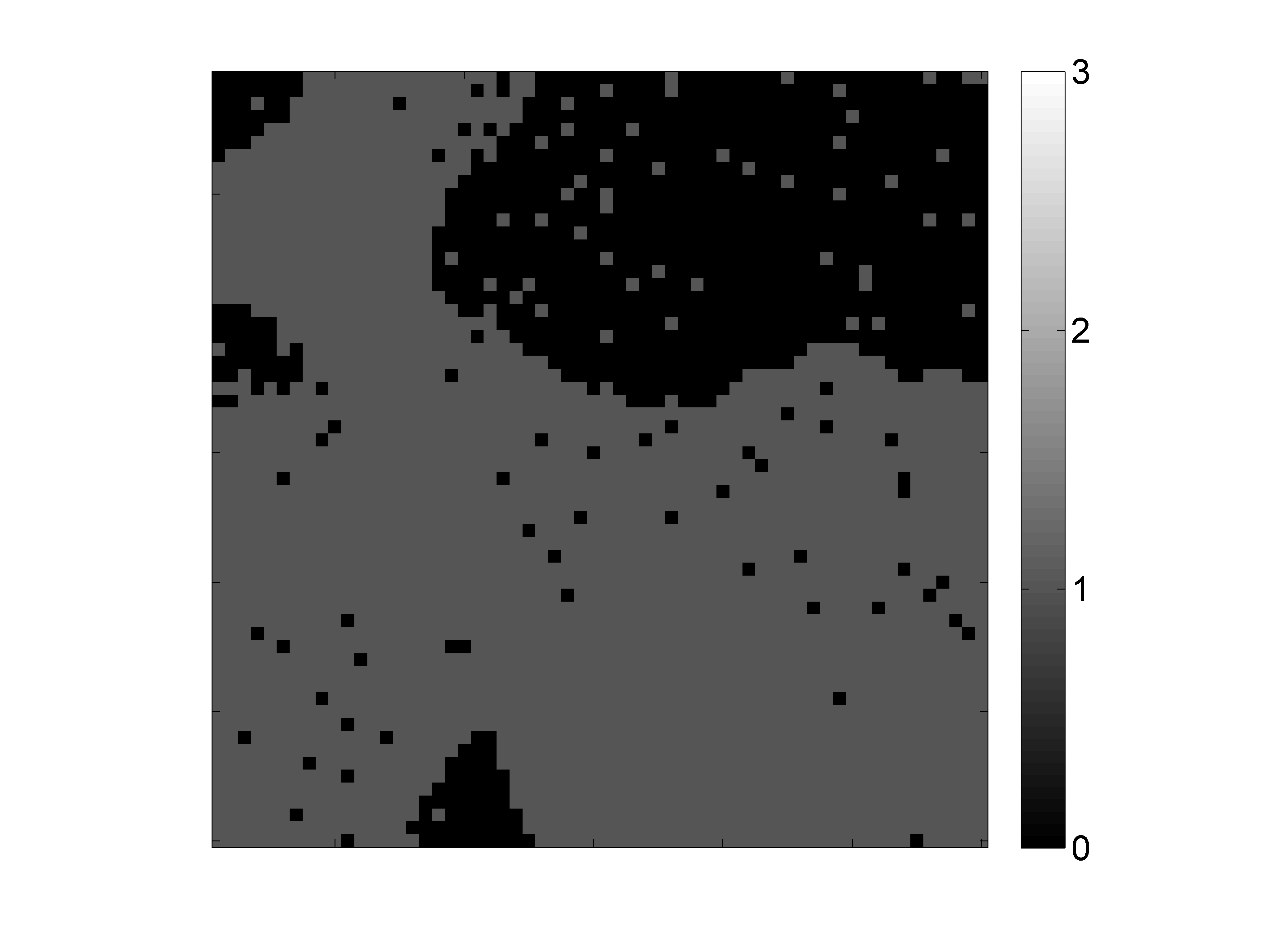}
 % \vspace{-0.2cm}
  \centerline{(c) Detection map (PPNMM).}\medskip
\end{minipage}
\hfill
\begin{minipage}[b]{0.48\linewidth}
  \centering
% \centerline{\epsfig{figure=image4.ps,width=4.0cm}}
\includegraphics[width=4.8cm]{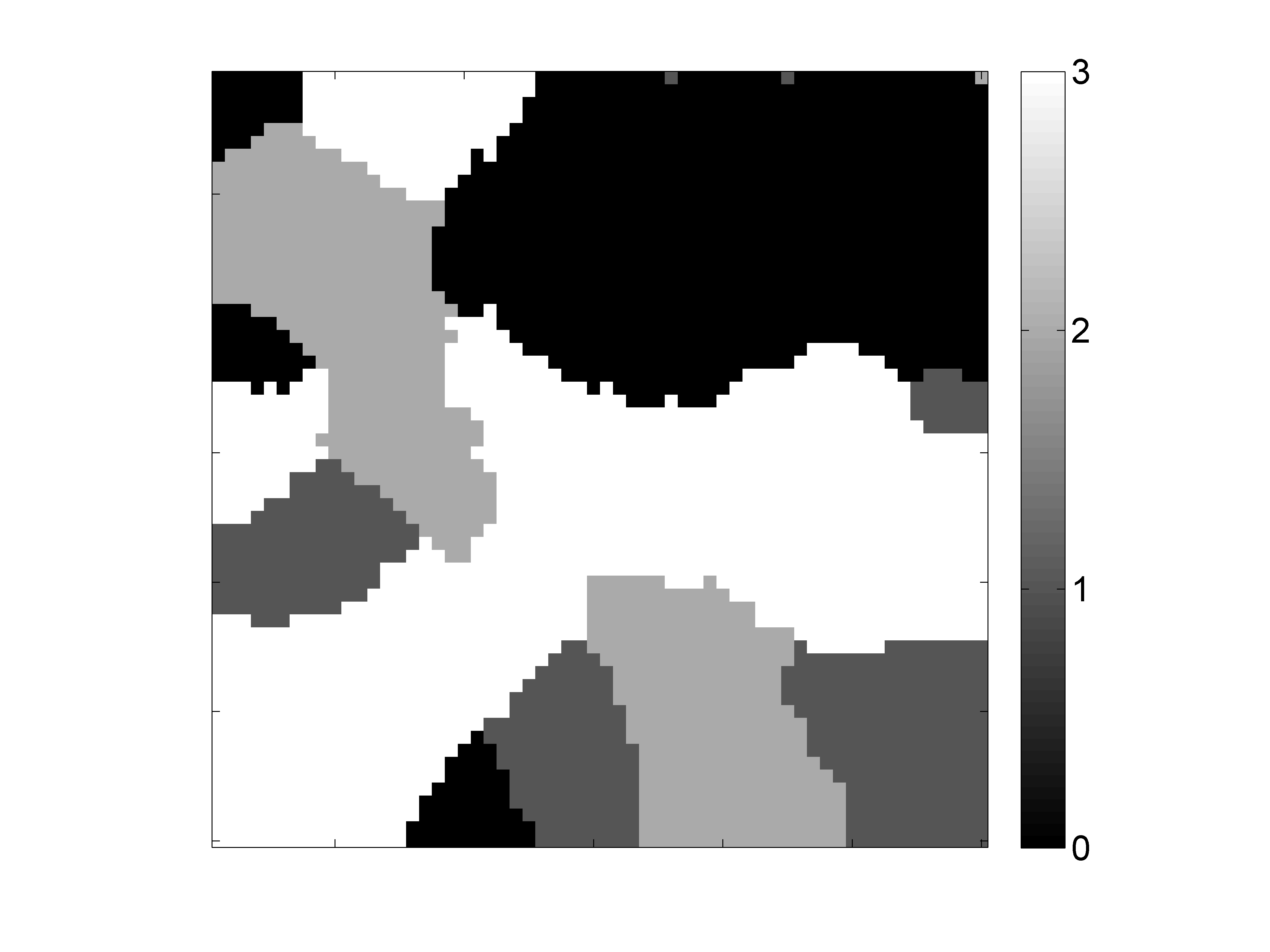}
 % \vspace{-0.2cm}
  \centerline{(d) Detection map (RCA-SU).}\medskip
\end{minipage}
\vspace{-0.3cm} \caption{Nonlinearity detection for the scenario \#2.} \label{fig:detection_4models_synth}
\end{figure}

The performance of the proposed joint nonlinear SU and nonlinearity
detection algorithm is then evaluated on a second synthetic image of
$60 \times 60$ pixels containing the $R=3$ spectral components
presented in the previous section. In this scenario, the image
consists of pixels generated according to four different mixing
models associated with four classes ($K=4$). The label map generated
using $\beta = 1.2$ is shown in Fig.
\ref{fig:detection_4models_synth} (a). The class $\calC_0$ is
associated with the LMM. The pixels of class $\calC_1$ have been
generated according to the generalized bilinear mixing model (GBM)
\cite{Halimi2010}
\begin{eqnarray}
\Vpix{n} & = & \sum_{r=1}^{R} \abond{r}{n}\Vmat{r}\nonumber\\
& + & \sum_{i=1}^{R-1}\sum_{j=i+1}^{R}\gamma_{i,j}\abond{i}{n}\abond{j}{n}\Vmat{i}\odot\Vmat{j} + \Vnoise_{\nopix}
\end{eqnarray}
where $n \in \mathcal{I}_1$ and the nonlinearity parameters
$\{\gamma_{i,j}\}$ have been uniformly drawn in $[0.5,1]$. The class
$\calC_2$ is composed of pixels generated according to the PPNMM
\cite{Altmann2012a} as follows
\begin{eqnarray}
\Vpix{n} & = & \sum_{r=1}^{R} \abond{r}{n}\Vmat{r}\nonumber\\
& + & b \left(\sum_{r=1}^{R} \abond{r}{n}\Vmat{r}\right)\odot\left( \sum_{r=1}^{R} \abond{r}{n}\Vmat{r}\right) + \Vnoise_{\nopix}
\end{eqnarray}
where $n \in \mathcal{I}_2$ and $b=0.5$ for all pixels in class $\calC_2$. Finally, the class
$\calC_3$ has been generated according to \eqref{eq:NLM0} with
$s^2=0.1$. For the four classes, the abundance vectors have been
randomly generated according to a uniform distribution over the
admissible set defined by the positivity and sum-to-one constraints.
All pixels have been corrupted by an additive i.i.d Gaussian noise of variance
$\noisevar = 10^{-4}$, corresponding to an average signal-to-noise
ratio $\mathrm{SNR} \simeq 30$dB. The noise is assumed to be i.i.d.
for a fair comparison with SU algorithms assuming
i.i.d. Gaussian noise. Fig. \ref{fig:detection_4models_synth} (b)
shows the log-energy of the nonlinearity parameters for each pixel
of the image, i.e., $\log\left(\norm{\bphi_n}^2\right)$ for $n=1,\ldots,3600$. This figure shows that each class corresponds to a different level of nonlinearity.
\subsubsection{Unmixing}
Different estimation procedures have been considered for the four
different mixing models:
\begin{itemize}
\item The FCLS
    algorithm \cite{Heinz2001} which is known to have good performance for linear mixtures.
\item The GBM-based approach \cite{Halimi2011IGARSS} which is particularly adapted for bilinear nonlinearities.
\item The
    gradient-based approach of \cite{Altmann2012a} which is based on a PPNMM and has shown nice properties for various nonlinear models.
\item The proposed RCA-SU algorithm which has been designed for the model in \eqref{eq:NLM0}. It has been applied to this data set
    with $N_{\textrm{MC}}=3000$, $N_{\textrm{bi}}=2000$, $K=4$
    and $\beta=1.2$.
\item Finally, we consider the K-Hype method \cite{Chen2012} to
    compare our algorithm with state-of-the art kernel based
    unmixing methods. The kernel used in this paper is the
    polynomial, second order symmetric kernel whose Gram matrix
    is defined by \eqref{eq:bK}. This kernel provides better
    performance on this data set than the kernels studied in
    \cite{Chen2012} (namely the Gaussian and the polynomial,
    second order asymmetric kernels). All hyperparameters of the
    K-Hype algorithm have been optimized using preliminary runs.
\end{itemize}

Table \ref{tab:RNMSE_synth2} compares the RNMSEs obtained with the
SU algorithms for each class of the second scenario. These
results show that the proposed algorithm provides abundance
estimates similar to those obtained with the LMM-based algorithm
(FCLS) for linearly mixed pixels. Moreover, the RCA-SU also provides
accurate estimates for the three mixing models considered, which
illustrates the robustness of the RCA-based model regarding model
mis-specification.

\begin{table}[h!]
\renewcommand{\arraystretch}{1.2}
\begin{footnotesize}
\begin{center}
\caption{Abundance RNMSEs ($\times 10^{-2}$): Scenario \#2
.\label{tab:RNMSE_synth2}}
\begin{tabular}{|c|c|c|c|c|}
\hline
\multirow{2}*{Unmixing algo.} &  Class \#0  & Class \#1 & Class \#2 & Class \#3  \\
  &  (LMM)  & (GBM)&  (PPNMM) & (RCA) \\
\hline
\multicolumn{1}{|c|}{FCLS} & $\textbf{\blue{0.35}}$ & $9.20$ & $19.74$ & $30.73$\\
\hline
\multicolumn{1}{|c|}{GBM} & $\textbf{0.36}$ & $3.05$ & $15.24$ & $29.53$\\
\hline
\multicolumn{1}{|c|}{PPNMM} & $0.65$ & $\textbf{\blue{1.37}}$ & $\textbf{\blue{0.48}}$ & $23.77$\\
\hline
K-HYPE  & $3.24$ & $3.28$ & $3.14$ & $\textbf{3.42}$\\
\hline
\multicolumn{1}{|c|}{RCA-SU} & $\textbf{\blue{0.35}}$ & $\textbf{1.58}$ & $\textbf{2.14}$ & $\textbf{\blue{3.41}}$\\
\hline
\end{tabular}
\end{center}
\end{footnotesize}
\vspace{-0.4cm}
\end{table}

The unmixing quality is also evaluated by the reconstruction error
(RE) defined as
\begin{eqnarray}
\label{eq:RE}
    \textrm{RE}_k= \sqrt{\dfrac{1}{N_k \nbband}\sum_{n \in \mathcal{I}_k}
    {\norm{\hat{\Vpixels}_{\nopix} - \Vpix{\nopix}}^2}}
\end{eqnarray}
where $\Vpix{\nopix}$ is the $\nopix$th observation vector and
$\hat{\Vpixels}_{\nopix}$ its estimate. Table \ref{tab:RE_synth2}
compares the REs obtained for the different classes. This table
shows the accuracy of the proposed model for fitting the
observations. The REs obtained with the RCA-SU are similar for the
four pixel classes. Moreover, the performance in terms of RE of the
proposed algorithm are similar to the performance of the K-Hype
algorithm.

\begin{table}[h!]
\renewcommand{\arraystretch}{1.2}
\begin{footnotesize}
\begin{center}
\caption{REs ($\times 10^{-2}$): Scenario \#2.\label{tab:RE_synth2}}
\begin{tabular}{|c|c|c|c|c|}
\hline
\multirow{2}*{Unmixing algo.} &  Class \#0  & Class \#1 & Class \#2 & Class \#3  \\
  &  (LMM)  & (GBM)&  (PPNMM) & (RCA) \\
\hline
\multicolumn{1}{|c|}{FCLS} & $\textbf{0.99}$ & $2.17$ & $1.33$ & $\textbf{3.10}$\\
\hline
\multicolumn{1}{|c|}{GBM} & $1.00$ & $1.12$ & $4.41$ & $10.98$\\
\hline
\multicolumn{1}{|c|}{PPNMM} & $\textbf{0.99}$ & $\textbf{1.01}$ & $\textbf{0.99}$ & $3.80$\\
\hline
  K-HYPE & $\textbf{\blue{0.98}}$ & $\textbf{\blue{0.98}}$ & $\textbf{\blue{0.98}}$ & $\textbf{\blue{0.98}}$\\
\hline
\multicolumn{1}{|c|}{RCA-SU} & $1.00$ & $\textbf{\blue{0.98}}$ & $\textbf{\blue{0.98}}$ & $\textbf{\blue{0.98}}$\\
\hline
\end{tabular}
\end{center}
\end{footnotesize}
\vspace{-0.4cm}
\end{table}
From a reconstruction point of view, the K-Hype and RCA-SU algorithms provides similar results. However, the proposed algorithm also provides nonlinearity detection maps. The PPNMM and RCA-SU algorithms perform similarly in term of abundance estimation and allow both nonlinearities to be detected in each pixel. However, the nonlinearities can be analyzed more deeply using the RCA-SU, as will be shown in the next part.
\subsubsection{Nonlinearity detection}
The performance of the proposed algorithm for nonlinearity detection
is compared to the detector studied in \cite{Altmann2012b}, which is
coupled with the PPNMM-based SU procedure mentioned above. The
probability of false alarm of the PPNMM-based detection has been set
to $\textrm{PFA}=0.05$. Figs. \ref{fig:detection_4models_synth} (c)
and (d) show the detection maps obtained with the two detectors.
Both detectors are able to locate the nonlinearly mixed regions.
However, the RCA-SU provides more homogeneous regions, due to
the consideration of spatial structure through the MRF. Moreover,
the proposed algorithm provides information about the different levels
of nonlinearity in the image thanks to the estimation of the hyperparameters $s_k^2$ associated with the different classes. In this simulation, we obtain
$[\hat{s}_1^2,\hat{s}_2^2,\hat{s}_3^2]= [0.2,1.4,10]\times
10^{-2}$, showing that nonlinearities
of class $\calC_1$ are less severe than those of class
$\calC_2$ and that are themselves weaker than those of class $\calC_3$. The next section studies the
performance of the proposed algorithm for a real hyperspectral
image.

\section{Simulations for a real hyperspectral image}
\label{sec:simu_real}
\subsection{Data set} 
\begin{figure}[h!]
  \centering
  \includegraphics[width=\columnwidth]{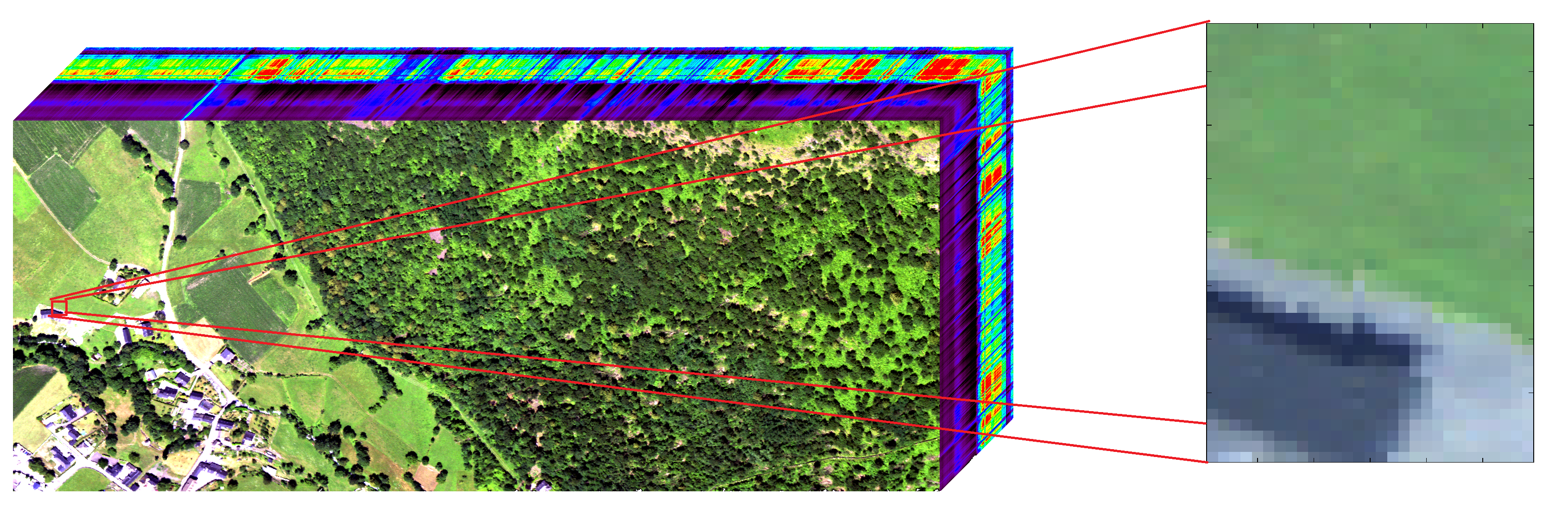}
  \caption{Real hyperspectral Madonna data acquired by the Hyspex hyperspectral scanner
   over Villelongue, France (left) and sub-image of interest (right).}
  \label{fig:Madonna_big}
\end{figure}

\begin{figure}[h!]
  \centering
  \includegraphics[width=\columnwidth]{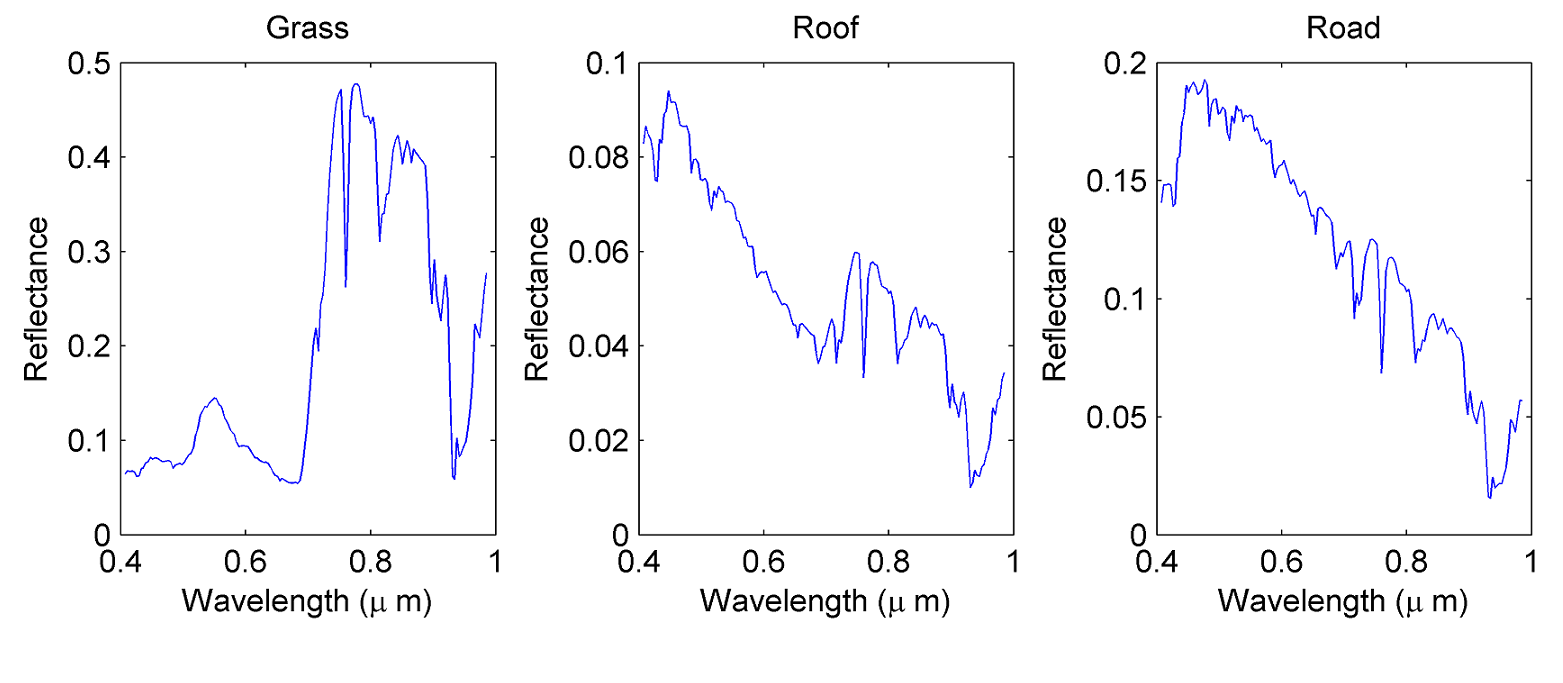}
  \caption{The $R=3$ endmembers estimated by N-Findr for the real Madonna sub-image.}
  \label{fig:Madonna_endmembers}
\end{figure}

The real image considered in this section was
acquired in 2010 by the Hyspex hyperspectral scanner over
Villelongue, France (00�03'W and 42�57'N). $L = 160$ spectral bands
were recorded from the visible to near infrared with a spatial
resolution of $0.5$m. This dataset has already been studied in
\cite{Sheeren2011,Altmann2013} and is mainly composed of forested
and urban areas. More details about the data acquisition and
pre-processing steps are available in \cite{Sheeren2011}. A
sub-image (of size $41 \times 29$ pixels) is chosen here to evaluate
the proposed unmixing procedure and is depicted in Fig.
\ref{fig:Madonna_big}. The scene is composed mainly of roof, road
and grass pixels, resulting in $R=3$ endmembers. The spectral
signatures of these components have been extracted from the data
using the N-FINDR algorithm \cite{Winter1999} and are depicted in
Fig. \ref{fig:Madonna_endmembers}.

\subsection{Spectral unmixing}
\begin{figure}[h!]
  \centering
  \includegraphics[width=0.7\columnwidth]{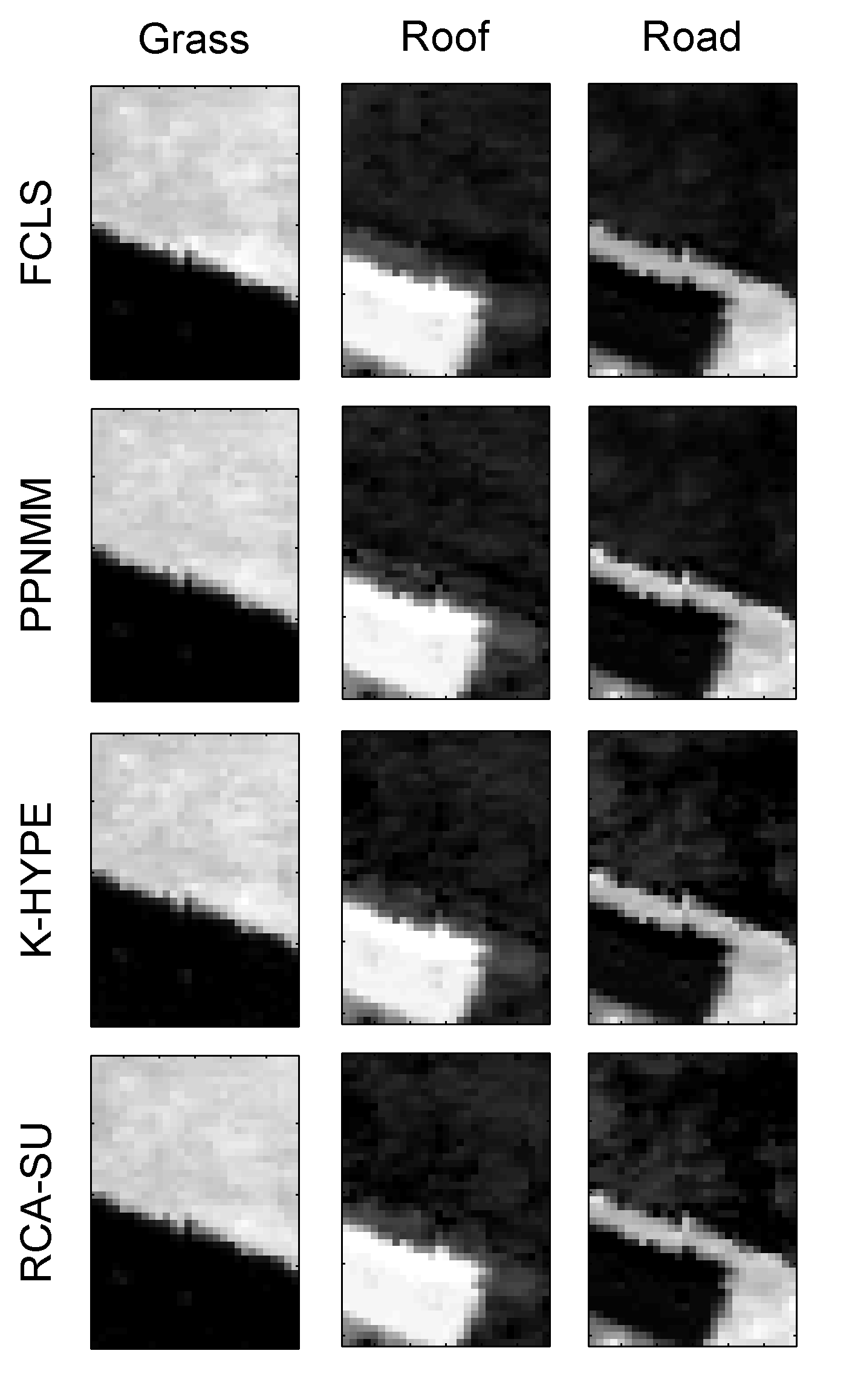}
  \caption{The $R=3$ abundance maps estimated by the FCLS, PPNMM-based, K-Hype, and RCA-SU algorithms for the Madonna real image (white pixels correspond to large abundances, contrary to black pixels).}
  \label{fig:Madonna_abundances}
\end{figure}

\begin{figure}[h!]
  \centering
  \includegraphics[width=\columnwidth]{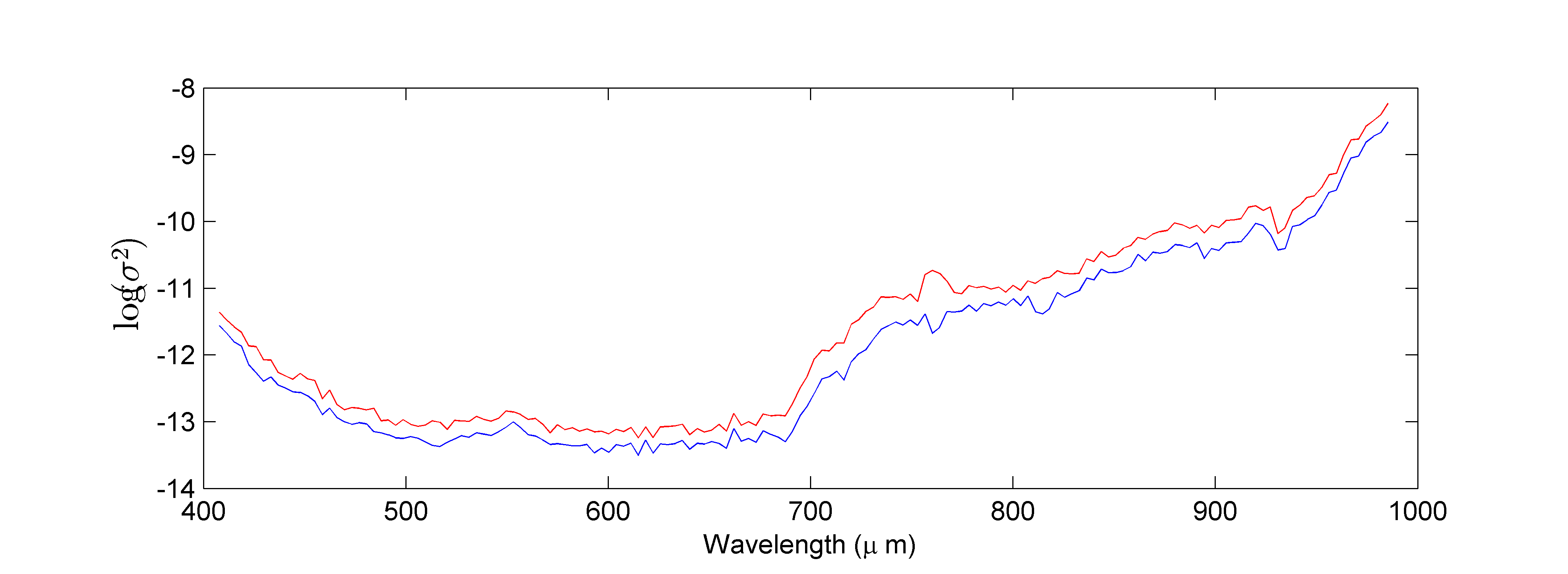}
  \caption{Noise variances estimated by the RCA-SU (red) and the Hysime
algorithm (blue) for the real Madonna image.}
  \label{fig:Madonna_variances}
\end{figure}

The proposed algorithm has been applied to this data set with
$N_{\textrm{MC}}=3000$ and $N_{\textrm{bi}}=1000$. The number of
classes has been set to $K=4$ (one linear class and three nonlinear classes). The
granularity parameter of the prior \eqref{eq:MRF_prior} has been
fixed to $\beta=0.7$. Fig. \ref{fig:Madonna_abundances} shows
examples of abundance maps estimated by the FCLS algorithm, the
gradient-based method assuming the GBM \cite{Halimi2011IGARSS},
the PPNMM \cite{Altmann2012a}, the K-Hype \cite{Chen2012}
algorithms and the proposed method. The abundance maps estimated by the RCA-SU algorithm are
in good agreement with the state-of-the art algorithms. However,
Table \ref{tab:RE_real} shows that K-Hype and the proposed algorithm provide a
lower reconstruction error. Fig. \ref{fig:Madonna_variances}
compares the noise variances estimated by the RCA-SU for the real
image with the noise variances estimated by the HySime algorithm
\cite{Bioucas2008}. The HySime algorithm assumes additive noise and
estimates the noise covariance matrix of the image using multiple
regression. Fig. \ref{fig:Madonna_variances} shows that the
two algorithms provide similar noise variance estimates. These results motivate the consideration of non i.i.d. noise for
hyperspectral image analysis since the noise variances increase for
the highest wavelengths. The simulations conducted on this real dataset show the accuracy of the proposed RCA-SU in terms of abundance estimation and reconstruction error, especially for applications where the noise variances vary depending on the wavelength. Moreover, it also provides information about the nonlinearities of the scene.

\begin{table}[h!]
\renewcommand{\arraystretch}{1.2}
\begin{footnotesize}
\begin{center}
\caption{Reconstruction errors: Real image.\label{tab:RE_real}}
\begin{tabular}{|c|c|}
\hline
\multirow{1}*{Unmixing algo.} &  RE ($\times 10^{-2}$) \\
\hline
\multicolumn{1}{|c|}{FCLS} & $0.65$ \\
\hline
\multicolumn{1}{|c|}{GBM} & $0.65$ \\
\hline
\multicolumn{1}{|c|}{PPNMM} & $\textbf{0.54}$\\
\hline
\multicolumn{1}{|c|}{K-HYPE} & $\textbf{\blue{0.48}}$\\
\hline
\multicolumn{1}{|c|}{RCA-SU} & $\textbf{\blue{0.48}}$ \\
\hline
\end{tabular}
\end{center}
\end{footnotesize}
\vspace{-0.4cm}
\end{table}

\subsection{Nonlinearity detection}
\begin{figure}[h!]
\begin{minipage}[b]{.48\linewidth}
  \centering
% \centerline{\epsfig{figure=image3.ps,width=4.0cm}}
\includegraphics[width=4cm]{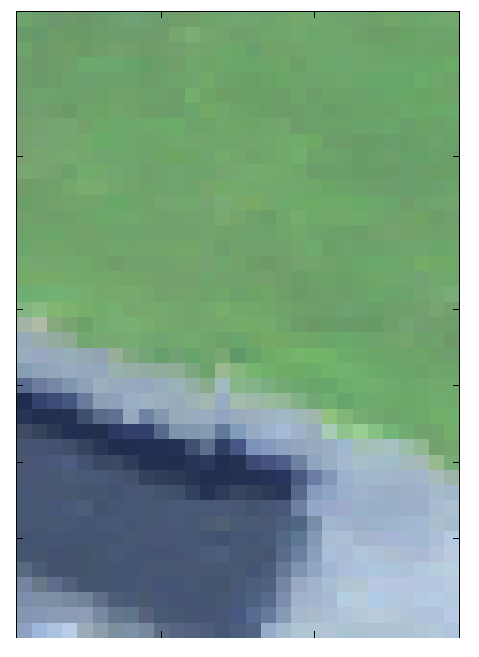}
 % \vspace{-0.2cm}
  \centerline{(a)}\medskip
\end{minipage}
\hfill
\begin{minipage}[b]{0.48\linewidth}
  \centering
% \centerline{\epsfig{figure=image4.ps,width=4.0cm}}
\includegraphics[width=4.8cm]{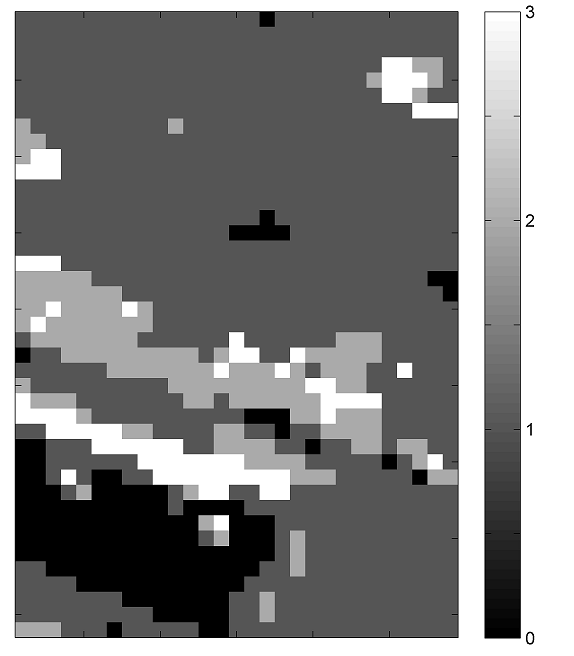}
 % \vspace{-0.2cm}
  \centerline{(b)}\medskip
\end{minipage}
\vspace{-0.3cm} \caption{(a) True color image of the scene of interest. (b) Nonlinearity detection map obtained with the RCA-SU detector for the Madonna image.} \label{fig:detection_Madonna}
\end{figure}

Fig. \ref{fig:detection_Madonna} (b) shows the detection map (map of $z_n$ for $n=1,\ldots,N$) provided by the proposed RCA-SU detector for the real image considered. Due to the consideration of spatial structures, the proposed detector provides homogeneous regions. Similar structures can be identified in this detection map and the true color image of the scene (Fig. \ref{fig:detection_Madonna} (a)). The estimated class $\calC_0$ (black pixels) associated with linearly mixed pixels is mainly located in the roof region. The class $\calC_1$ (dark grey pixels) can be related to regions where the main component in the pixels are grass or road. Mixed pixels composed of grass and road are gathered in class $\calC_2$ (light grey pixels). Finally, shadowed pixels located between the roof and the road are associated with the last class $\calC_3$ (white pixels).  Moreover, the RCA-SU can identify three levels of nonlinearity, corresponding to $[\hat{s}_1^2,\hat{s}_2^2,\hat{s}_3^2]=[0.03,0.50,29.5]$. The most influent nonlinearity class is class $\calC_3$, where shadowing effects occurs. Mixed pixels of class $\calC_2$ contain weaker nonlinearities. Finally, the remaining pixels of class $\calC_1$ are associated with the weakest nonlinearities. The nonlinearities of this class can probably be explained by the endmember variability and/or the endmember estimation error. It is interesting to note that the RCA-SU identifies two rather linear classes associated with homogeneous regions mainly composed of a single parameter (classes $\calC_0$ and $\calC_1$). The two latter classes (classes $\calC_2$ and $\calC_3$) correspond to rather nonlinear regions where the pixels are mixed and shadowing effects occur.
\section{Conclusion}
\label{sec:conclusion} We have proposed a new hierarchical Bayesian algorithm for joint linear/nonlinear spectral unmixing of hyperspectral images and nonlinearity detection. This algorithm assumed that each pixel of the image is a linear or nonlinear mixture of endmembers contaminated by additive Gaussian noise. The nonlinear mixtures are decomposed into a linear combination of the endmembers and an additive term representing the nonlinear effects. A Markov random field was introduced to promote spatial structures in the image. The image was decomposed into regions or classes where the nonlinearities share the same statistical properties, each class being associated with a level of nonlinearity. Nonlinearities within a same class were modeled using a Gaussian process parameterized by the endmembers and the nonlinearity level. Note finally that the physical constraints for the abundances were included in the Bayesian framework through appropriate prior distributions. Due to the complexity of the resulting joint posterior distribution, a Markov chain Monte Carlo method was investigated to compute Bayesian estimators of the unknown model parameters.

Simulations conducted on synthetic data illustrated the performance of the proposed algorithm for linear and nonlinear spectral unmixing. An important advantage of the proposed algorithm is its robustness regarding the actual underlying mixing model. Another interesting property resulting from the nonlinear mixing model considered is the possibility of detecting several kinds of linearly and nonlinearly mixed pixels. This detection can be used to identify the image regions affected by nonlinearities in order to characterize the nonlinear effects more deeply. Finally, simulations conducted with real data showed the accuracy of the proposed unmixing and nonlinearity detection strategy for the analysis of real hyperspectral images.

The endmembers contained in the hyperspectral image were
assumed to be known in this work. Of course, the performance of the
algorithm relies on this endmember knowledge. We think that
estimating the pure component spectra present in the image, jointly
with the abundance estimation and the nonlinearity detection is an
important issue that should be considered in future work. Finally,
the number of classes and the granularity of the scene were assumed to
be known in this study. Estimating these parameters is clearly a
challenging issue that is under investigation.

\bibliographystyle{IEEEtran}
\bibliography{biblio}
\end{document}